\documentclass{ws-ijmpe}
\usepackage[super,compress]{cite}
\usepackage{xspace}
\usepackage[driverfallback=dvipdfm]{hyperref}

\usepackage{color}

\definecolor{dgreen}{cmyk}{1.,0.,1.,0.2} % dark green
\definecolor{orange}{cmyk}{0.,0.353,1.,0.} % orange

\newcommand{ \be }{\begin{equation}}
\newcommand{ \ee }{\end{equation} }
\newcommand{ \bea }{\begin{eqnarray}}
\newcommand{ \eea }{\end{eqnarray} }

\newcommand{ \ds }{\displaystyle}

\newcommand{ \bA }{{\bf A}}
\newcommand{ \bB }{{\bf B}}
\newcommand{ \bs }{{\bf s}}
\newcommand{ \bS }{{\bf S}}
\newcommand{ \bl }{{\bf l}}
\newcommand{ \bv }{{\bf v}}
\newcommand{ \bP }{{\bf P}}
\newcommand{ \bp }{{\bf p}}

\def \xim {\Xi^-}

\newcommand{ \mean }[1]{\left\langle #1 \right\rangle}

\newcommand{\snn}{\ensuremath{\sqrt{s_{\rm NN}}}\xspace}

\newcommand{\pz}{\ensuremath{P_{\rm z}}\xspace}
\newcommand{\px}{\ensuremath{P_{\rm x}}\xspace}
\newcommand{\py}{\ensuremath{P_{\rm y}}\xspace}
\newcommand{\pmy}{\ensuremath{P_{\rm -y}}\xspace}

\newcommand{\PsiRP}{\ensuremath{\Psi_{\rm RP}}\xspace}

\newcommand{\psirp}{\ensuremath{\Psi_{\rm RP}}\xspace}

\newcommand{\phist}{\ensuremath{\phi^\ast}\xspace}

\newcommand{\pt}{\ensuremath{p_{\rm T}}\xspace}

\newcommand{ \pd }{{\partial}}

\newcommand{\lam}{\ensuremath{\Lambda}\xspace}
\newcommand{\alam}{\ensuremath{\bar{\Lambda}}\xspace}
\newcommand{\xin}{\ensuremath{\Xi^{-}}\xspace}
\newcommand{\xibar}{\ensuremath{\bar{\Xi}^{+}}\xspace}
\newcommand{\om}{\ensuremath{\Omega^{-}}\xspace}
\newcommand{\ombar}{\ensuremath{\bar{\Omega}^{+}}\xspace}

\newcommand{\ximunu}{\xi^{\mu\nu}}
\newcommand{\dnu}{\partial^{\nu}}
\newcommand{\dmu}{\partial^{\mu}}
\newcommand{\unu}{u^{\nu}}
\newcommand{\umu}{u^{\mu}}

\newcommand{ \bomega }{{\boldsymbol \omega}}

\newcommand{\tomega}{\varpi}

\newcommand{ \grad }{{\boldsymbol \nabla}}

\begin{document}

\markboth{T. Niida and S. Voloshin}{Polarization in heavy ion collisions}

\catchline{}{}{}{}{}
%%%%%%%%%%%%%%%%%%%%%%%%%%%%%%%%%%%%%%%%%%%%%%%%%%%%%%%%%%%%%%%%%%%%

\title{Polarization phenomenon in heavy-ion collisions 
  %\footnote{For the title, try not to use more than 3 lines.
  %Typeset the title in 10 pt Times roman, uppercase and boldface.}
}

\author{Takafumi Niida}
%\footnote{Typeset names in 10~pt Times roman, uppercase. Use the
%footnote to indicate the present or permanent address of the author.}
\address{Department of Physics, University of Tsukuba, 1-1-1 Tennodai\\
  Tsukuba, Ibaraki 305-8571, JAPAN\\
  niida.takafumi.fw@u.tsukuba.ac.jp}
%\footnote{State completely without abbreviations, the affiliation and
%mailing address, including country. Typeset in 8~ptTimes italic.}\\

\author{Sergei A. Voloshin}

\address{Department of Physics and Astronomy, Wayne State University,
  666 W. Hancock\\
  Detroit, Michigan 48201, 
  USA\\
  sergei.voloshin@wayne.edu}

\maketitle

\begin{history}
%\received{30 November 2021}
\received{16 April 2023}
%\revised{Day Month Year}
%\revised{\new{version 6, \today}}
%\accepted{Day Month Year}
%\comby{(xxxxxxxxxx)}
\end{history}

\begin{abstract}
%The abstract should be summarized in less than 200 words. It should
%not contain any references or displayed equations. Typeset the
%abstract in 8~pt Times roman with baselineskip of 10~pt, making
%an indentation of 1.5 pica on the left and right 
The strongly interacting system created in ultrarelativistic nuclear
collisions behaves almost as an ideal fluid with rich patterns of the
velocity field exhibiting strong vortical structure. Vorticity of the
fluid, via spin-orbit coupling, leads to particle spin polarization.
Due to the finite orbital momentum of the system, the polarization on
average is not zero; it depends on the particle momenta reflecting
the spatial variation of the local vorticity.

In the last few years, this field experienced a rapid growth due to
experimental discoveries of the global and local polarizations. Recent
measurements triggered further development of the theoretical
description of the spin dynamics and suggestions of several new
mechanisms for particle polarization. In this review, we focus mostly
on the experimental results.  We compare the measurements with the
existing theoretical calculations but try to keep the discussion of
possible underlying physics at the qualitative level. Future
measurements and how they can help to answer open theoretical
questions are also discussed. We pay a special attention to the
employed experimental methods, as well as to the detector effects and
associated corrections to the measurements.
\end{abstract}


\ccode{PACS numbers:}

\tableofcontents

%=================================================================
%\section{Introduction} \label{sec:intro}
%============================================================
\section{Introduction: Polarization as a collective phenomenon}

The discovery of the global polarization in heavy-ion collisions, the
hyperon polarization along the system orbital
momentum~\cite{STAR:2017ckg,Adam:2018ivw}, followed by the
measurements of the polarization along the beam
direction~\cite{Adam:2019srw}, opened totally new opportunities for
study of the nuclear collision dynamics and the properties of the
quark-gluon plasma (QGP), as well as for deeper understanding of the
spin and its transport in QGP medium.  These polarization measurements
are among the most significant discoveries made in heavy-ion collision
program along with observations of the strong elliptic flow and jet
quenching~\cite{STAR:2000ekf,STAR:2002svs,PHENIX:2003qdj,Voloshin:2008dg},
and have generated intense theoretical discussions as well as
experimental activities.

The phenomenon of the global polarization in heavy-ion collisions
arises from the partial conversion of the orbital angular momentum of
the colliding nuclei into the spin angular momentum of produced
particles~\cite{Liang:2004ph,Voloshin:2004ha,Becattini:2007sr}.  As a
result, the particles on average become polarized along the direction
of the initial orbital momentum of the two colliding nuclei.  The term
``global'' in the name of the phenomenon indicates that the component
of the particle polarization along the system orbital momentum is not
zero when averaged over all produced particles. The origin of the
polarization in heavy ion collisions is in the collective motion of
the strongly interacting fluid, and it is unlikely to be related to
the hyperon polarization (with respect to the production plane)
observed in $pp$ and $pA$ collisions~\cite{Nurushev:2013vn}.

\begin{figure}[th]
\centerline{
  \includegraphics[width=0.8\linewidth]{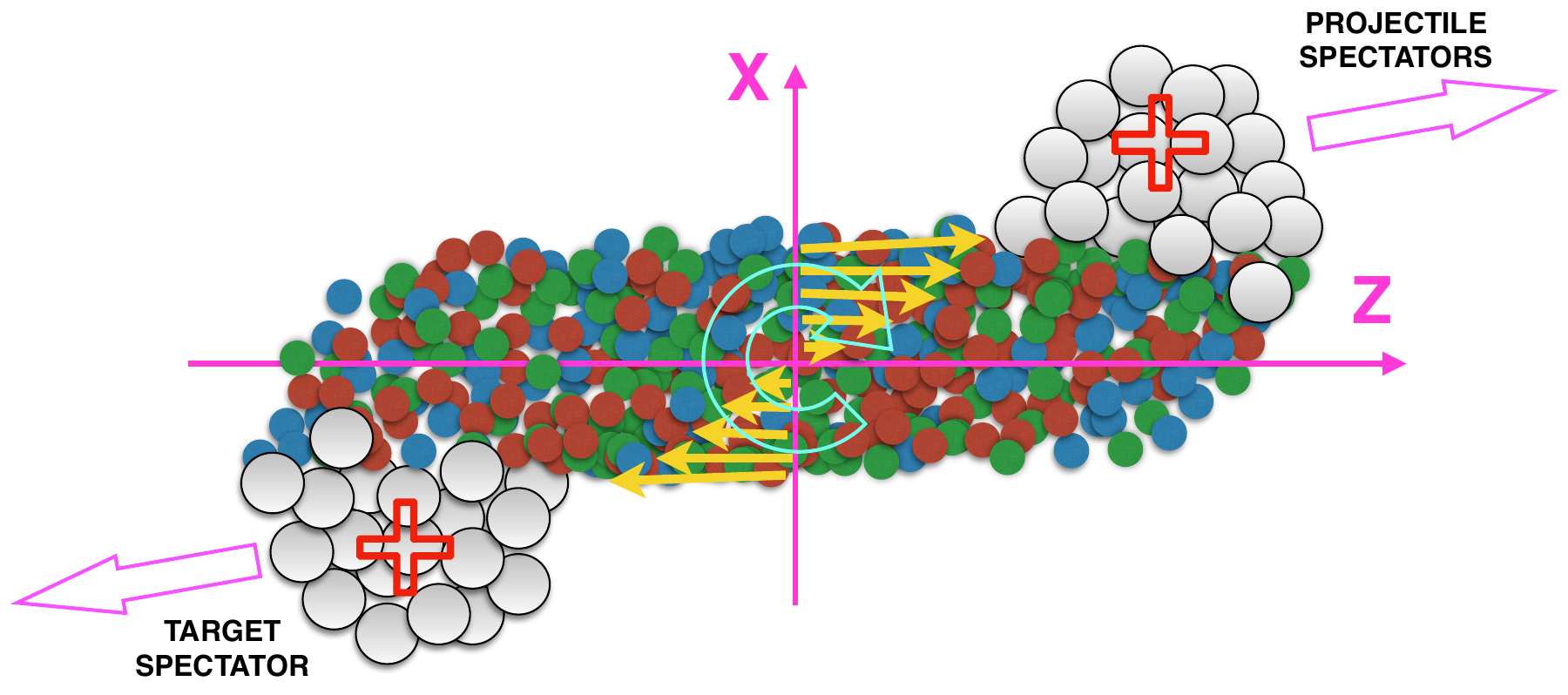}
}
\caption{Schematic view of a nuclear collision with $x$ and $z$ being
  the impact parameter and beam directions, respectively.  The system
  orbital angular momentum as well as the magnetic field points into
  the page, opposite to the direction of the $y$ axis.  Solid yellow
  arrows indicate collective velocity field at $z=0$. Open arrow
  indicates vorticity of the system.}
\label{fig:cartoon}
\end{figure}

In a non-central nuclear collision, the most prominent pattern in the
initial collective velocity distribution is a shear of the velocity
field, $dv_z/dx \ne 0$, where the $z$ direction is chosen along the
beam, and the $x$ direction is along the impact parameter vector
defined as a vector connecting the centers of the two nuclei (pointing
from the ``target'' to the ``projectile'' nucleus, the latter defined
as the one moving in the positive $z$ direction, see
Fig.~\ref{fig:cartoon}.  Such a shear in the velocity field leads to
nonzero vorticity characterizing the local orbital angular momentum
density. Particle binary interactions in the system would have on
average non-zero orbital angular momentum, which will be partially
converted into spin of the final-state particles. For example, in a
system with non-zero vorticity consisting of pions, the colliding
pions have a preferential direction for their orbital angular
momentum, and the spin of $\rho$ mesons produced in such collisions
($\pi^+\pi^-\rightarrow\rho^0$) would point in that very
direction~\cite{Voloshin:2004ha}.

\vspace*{-5mm}

The idea of the global polarization is almost 20 years old with the
initial predictions for the quark and final particles polarization as
high as ``in the order of tens of a percent''~\cite{Liang:2004ph}.  As
pointed in Ref.~\cite{Voloshin:2004ha}, the global polarization
phenomenon, if that strong, could affect the interpretation of
different measurements.  In particular, the polarization of the vector
mesons would have a significant contribution to the elliptic flow
measurements~\cite{Voloshin:2004ha,Liang:2004xn}.  The decay products
of the vector resonances with spin pointing perpendicular to the
reaction plane have the angular distribution enhancing the in-plane
particle production and thus contributing to the elliptic flow
measurements.  The first measurements~\cite{STAR:2007ccu} of the
global polarization of \lam hyperons in Au+Au collisions at 200~GeV by
the STAR Collaboration put an upper limit on hyperon polarization of
$|P_\Lambda|\le 0.02$. Subsequently, the theoretical predictions have
been improved~\cite{Gao:2007bc} to be consistent with these
experimental results.

Particle polarization is determined by the vorticity of the fluid
element where the particle has been produced.  Due to the strong
space-momentum correlation present in the system, the polarization of
the particles in a certain momentum range would reflect the vorticity
in the regions where those particles are predominantly emitted
from. Thus, while averaged over the entire system only the ``global''
polarization component survives, the polarization in general depends
on the particle momentum. Such polarization is often referred to as
``local''. The non-trivial local vorticity can originate, for example,
due to propagation of highly energetic jets produced by the
partons hard scattering~\cite{Betz:2007kg,Serenone:2021zef}, or
due to collective expansion of the
system~\cite{Pang:2016igs,Voloshin:2017kqp,Becattini:2017gcx,Xia:2018tes}.
An important example of that is the polarization along the
beam direction due to (transverse) anisotropic flow, discussed first
in Ref.~\cite{Voloshin:2017kqp} on a basis of a simple Blast-Wave 
model, as well as observed in full hydrodynamical
calculations~\cite{Becattini:2017gcx}.

A short review format does not allow to describe and discuss in detail
all the results and the questions under discussion.  Our goal is to
provide a more general picture of the field, to emphasize the major
developments in our understanding of the phenomena and formulate
outstanding questions, and to outline the relation of the current and
future measurements to the underlying physics. In the first section of
the review, we focus the discussion on the nature of the
phenomenon. We use a very simple picture based on Glauber and
Blast-Wave models for illustrations and rough estimates. Then we
discuss the experimental side of the measurements, emphasizing the
details important for the interpretation of the results and their
uncertainties, as well as what is needed to accomplish this or other
measurements. We proceed with an overview of available results and
their current theoretical interpretations.  The overview of the
experimental results is followed by a summary of what we have learned
so far, open questions, and future perspectives.

In the following discussion we refer to $x$, $y$, and $z$ components
of the polarization according to the coordinate system
depicted in Figs.~\ref{fig:cartoon} and \ref{fig:glauber}. In such a
coordinate system, the global polarization would be given by the
average of $-\py=\pmy$, and the polarization along the beam direction is
given by $\pz$.  The global {\em averages} (over all particles and momenta)
of $\pz$ and $\px$ components (more exactly, the components of the
total spin angular momentum) are expected to be zero.

  % 

%====================================

%=====================================================================
%\section{What to expect?}
\section{Global and local polarizations}
\label{sec:phenomenon}

%----------------------------------------------------------------------
\subsection{Nonrelativistic vorticity and %estimates of
  the global polarization, $\mean{\pmy}$ }
\label{sec:estimates}

A very significant development leading to a fast progress in this
field was an application of the statistical methods to vortical fluid
with non-zero spin particles~\cite{Becattini:2007nd}, and development
of the hydrodynamical calculations based on the assumption of the
local angular momentum equilibrium~\cite{Becattini:2013fla}.  A rough
estimate of the polarization can be obtained with the help of a simple
nonrelativistic~\footnote{Note that for hyperons used in polarization
measurements $m_H \gg T$, where $T$ is the temperature and $m_H$ is a
hyperon mass.} formula describing the particle distribution in a
fluid with nonzero vorticity~\cite{Becattini:2016gvu}:
\begin{equation}
w \propto \exp\left[-\left(E +\bomega (\bs+\bl) -\frac{\mu}{s} \, 
\bs \bB \right)/T\right],
\label{eq:nr}
\end{equation}  
where $\bomega=\frac{1}{2}\,\grad \times \bv$ is the local vorticity
of the fluid velocity ($\bv$) field, ${\mathbf B}$ is an external
magnetic field, $\mu$ is the particle magnetic moment, 
and $T$ is the temperature of the system at equilibrium.  
Then the polarization of particles with spin $s$ is given by:
\be
\bP=\frac{\mean{\bs}}{s} \approx \frac{(s+1)}{3}\frac{(\bomega+\mu \bB/s)}{T}.
\label{eq:polar}
\ee
For spin 1/2 particles, the vorticity contribution to the polarization
is $P=\omega/2T$~\footnote{The nonrelativistic estimate can be also
obtained by noting that the entropy $\sigma (E)$ of the rotating gas
can be approximated as $\sigma(E-L^2/(2I))$, where $L$ is the orbital
momentum, and $I$ is the system inertia. Under condition of angular
momentum conservation, $S+L=\rm const$, this leads to $\partial \sigma
/\partial S = L/(IT)=\omega/T$.}.  Averaged over the entire system
volume, the vorticity direction coincides with the direction of the
system orbital angular momentum.  Note that the magnetic field created
by fast positively charged nuclei is also pointing in the same
direction.

\begin{figure}[th]
\centerline{
\includegraphics[width=0.32\linewidth]{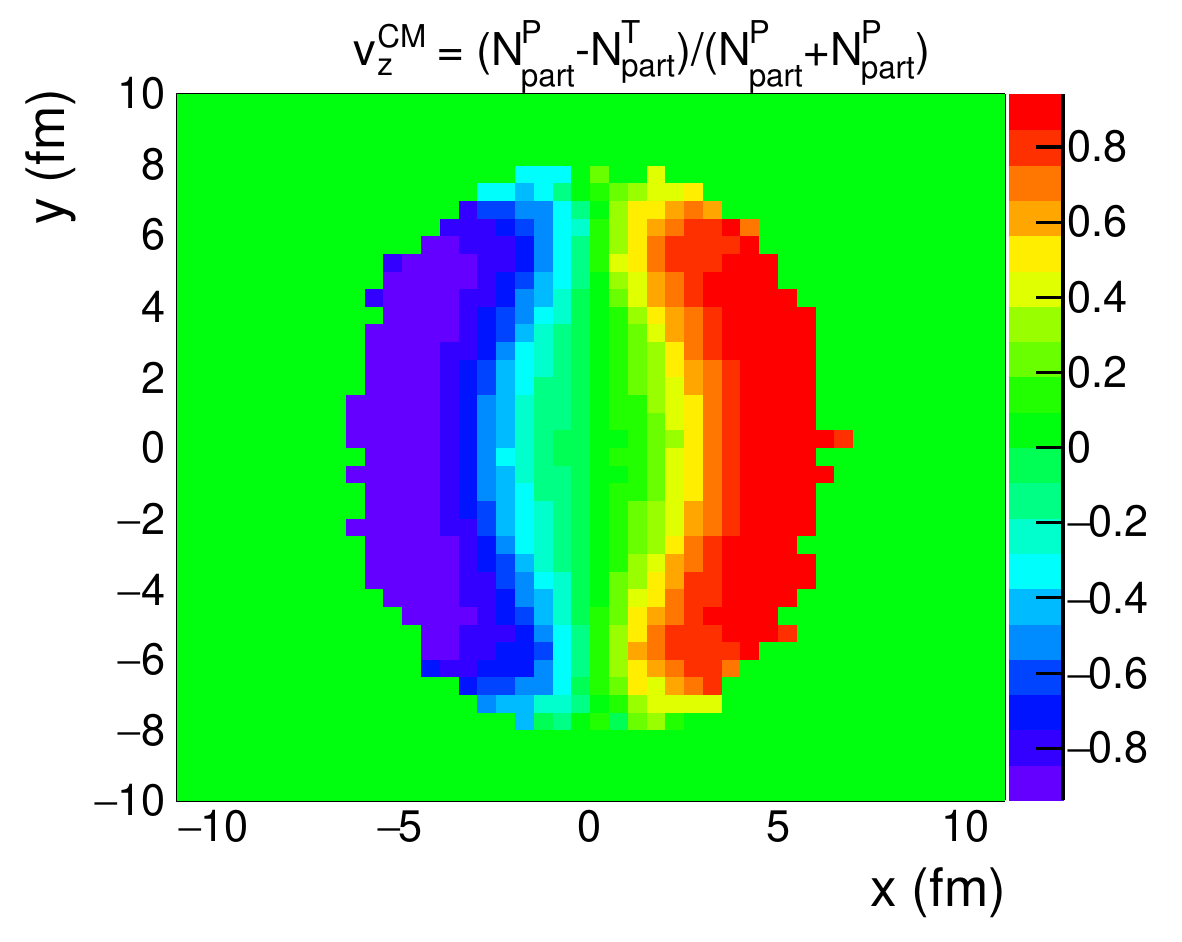}
\includegraphics[width=0.32\linewidth]{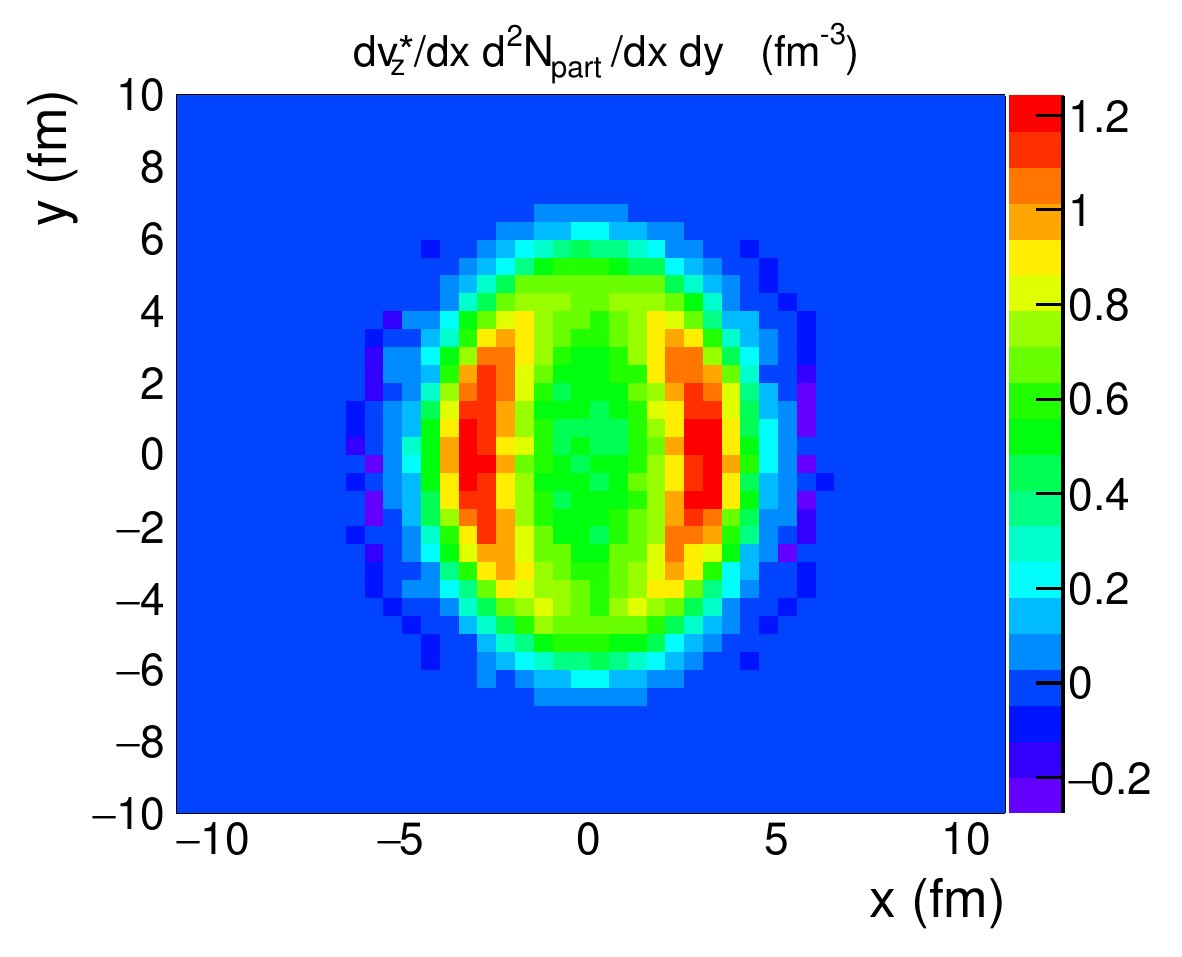}
\includegraphics[width=0.32\linewidth]{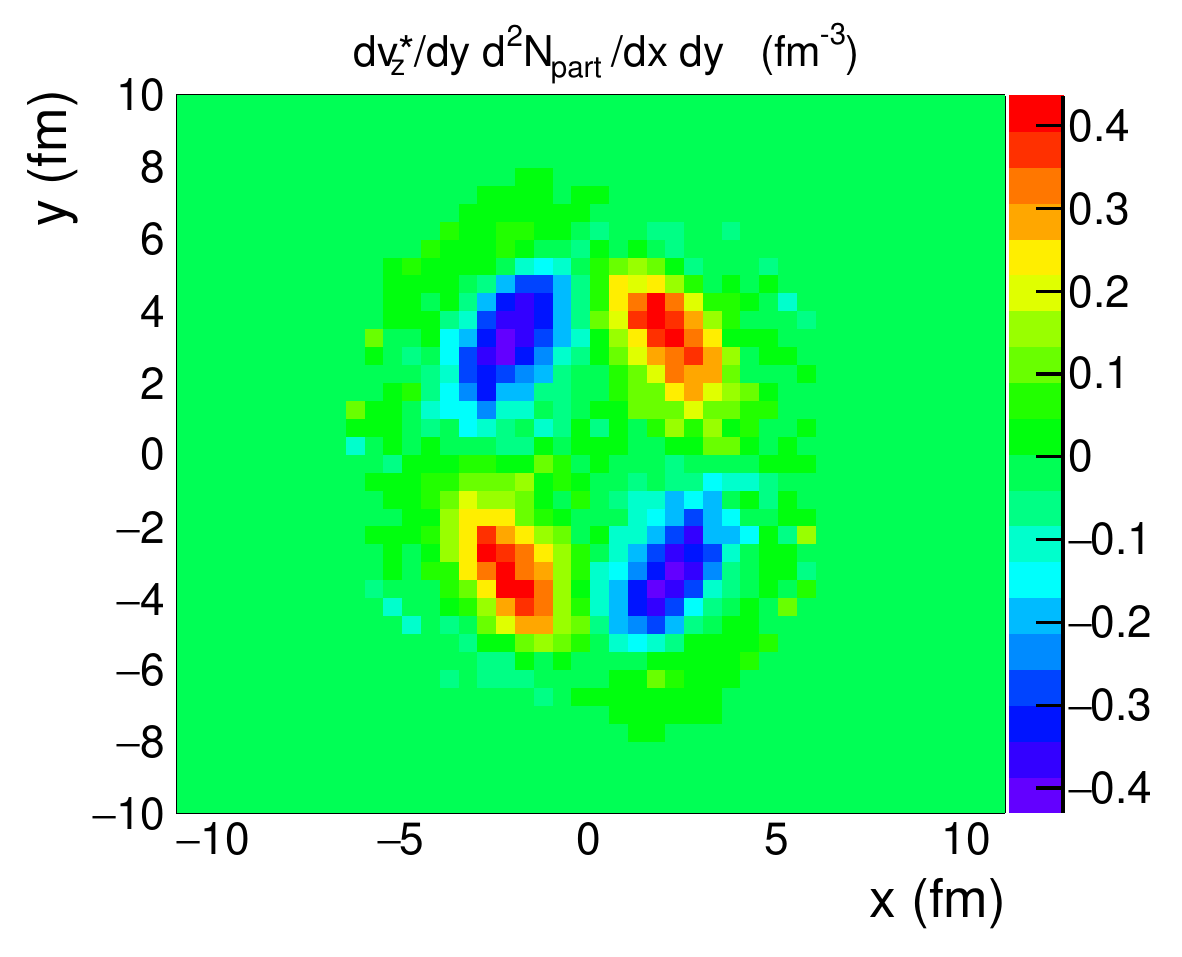}
}
\caption{ (a) A transverse plane distribution of the $z$-component of
  the velocity of participant nucleons in the center-of-mass frame,
  (b) $dv_z/dx$ distribution weighted with participant nucleon
  density, (c) the same for $dv_z/dy$, based on the Glauber model.  }
\label{fig:glauber}
\end{figure}

Figure~\ref{fig:cartoon} shows a cartoon of a non-central nuclear
collision with solid arrows indicating the velocity field of the
matter at the plane $z=0$. One can estimate the vorticity as $\omega_y
\approx -\frac{1}{2}\partial v_z/\partial x$ where $v_z$ is the
net-velocity along the beam direction that depends on the number of
participants coming from the target vs. projectile nuclei. The
magnitude of $v_z$ is reflected in the length of the solid arrows in
Fig.~\ref{fig:cartoon}.  For a rough estimate of the vorticity, we
present the velocity ($v_z$) distribution in the transverse plane in
Fig.~\ref{fig:glauber}(a).  In these calculations, the velocity was
estimated as $v_z=(n_{\rm P}-n_{\rm T})/(n_{\rm P}+n_{\rm T})$ where
$n_{\rm P}$ and $n_{\rm T}$ are the densities of the projectile and
target nucleon participants (nucleons that experienced inelastic
collisions) obtained by a simple Glauber model. The middle and right
plots in Fig.~\ref{fig:glauber} show the derivatives $dv_z^*/dx$ and
$dv_z^*/dy$ (with the asterisks denoting the quantities in the rest
frame of the fluid) weighted with the density of participating
nucleons (roughly proportional to the produced particle density).
From these estimates, one concludes that the vorticity might be as
large as a few percent of $\rm fm^{-1}$.  Then the nonrelativistic
formula~(\ref{eq:nr}) yields for the spin 1/2 particle polarization,
$P \approx \omega/(2T)$, in the range of a few percent (assuming
$T\sim 100$~MeV). Note that this simple estimate ignores the effect of
nuclear transparency at high energies where the vorticity values could
be significantly lower.

The vorticity of the system, especially its component along the
system's orbital momentum, is directly related to the asymmetries in
the initial velocity fields and thus it is intimately related to the
directed flow
$v_1$~\cite{Csernai:2013bqa,Bozek:2010bi,Becattini:2015ska}.  The
$v_1$ is defined by the first Fourier moment $v_1 = \langle \cos
(\varphi-\PsiRP)\rangle$ of the produced particles' azimuthal
distribution relative to the collision reaction plane angle \PsiRP.
Hydrodynamic simulations show that the orbital angular momentum stored
in the system and the directed flow of charged particles are almost
directly proportional to each other~\cite{Becattini:2015ska}.  This
allows for an empirical estimate of the collision energy dependence of
the global polarization~\cite{Voloshin:2017kqp}.
The STAR results for the directed
flow~\cite{STAR:2014clz,Singha:2016mna} and the hyperon global
polarization~\cite{STAR:2017ckg,Adam:2018ivw} from the RHIC Beam
Energy Scan program show that the slopes of $v_1$ at midrapidity
(${\rm d}v_1/{\rm d}\eta$) for charged hadrons and the hyperon
polarization are indeed strongly correlated.  The charged-particle
directed flow in Pb--Pb collisions at $\snn =
2.76$~TeV~\cite{ALICE:2013xri} is about three times smaller than at
the top RHIC energy of 200~GeV~\cite{STAR:2008jgm}.  This suggest that
the global polarization at the LHC energies should be also about three
times smaller than at RHIC %(around $\sim~0.08$\%)
and decreasing from
$\snn~=~2.76$~TeV to 5.02~TeV by about
$\sim20$\%~\cite{ALICE:2019sgg}.  Even smaller polarization values at
the LHC are expected when the directed flow is considered as a
combination of the two effects -- the tilt of the source in the
longitudinal direction and the dipole flow originating from the
asymmetry in the initial energy density
distributions~\cite{Adamczyk:2017ird}.  Taking into account that only
the contribution to the directed flow from the tilted source is
related to the vorticity and that its contribution relative to the
dipole flow decreases with the collision
energy~\cite{Adamczyk:2017ird}, one arrives to an estimate for the
global polarization at the LHC energies
of the order of 0.15--0.2 of that at the top RHIC energy.

%------------------------------------------------------------
\subsection{Role of the magnetic field}
\label{sec:Bfield}

In addition to possessing of the large orbital angular momentum, the
system also experiences a strong magnetic field, of the order of
$B\sim e/m_{\pi}^2 \sim 10^{14}$~T, generated in the initial state of
the collision~\cite{Kharzeev:2007jp,Skokov:2009qp,Voronyuk:2011jd} by
the fast moving electrically charged nuclei and by the spectator
protons after the collision.  The direction of the magnetic field
coincides with that of the orbital angular momentum.  Therefore, the
measured global polarization would include a contribution from the
magnetic field, see Eq.~\eqref{eq:polar}, especially if the magnetic
field is sustained for longer time by the presence of the
QGP~\cite{Tuchin:2013ie}.  Unlike to the case for vorticity
contribution, the contribution from the magnetic field is opposite for
particles and antiparticles because of the difference in signs of the
magnetic moments $\mu$ in Eq.~\eqref{eq:polar}.  The lifetime of the
initial magnetic field depends on the electric conductivity of the
QGP, which is poorly known.  Precise measurements of the polarization
difference between particles and antiparticles can provide an
important constraint on the magnitude of the magnetic field at the
hadronization time~\cite{Becattini:2016gvu,Muller:2018ibh}, as well as
on medium conductivity.  Such information is of particular importance
for study of the chiral magnetic effect~\cite{Kharzeev:2015znc}.

At lower collision energies, the passing time of two nuclei becomes
larger and therefore the lifetime of the magnetic field is
extended. Also, the medium created in the collision has positive
net-charge due to baryon stopping.  If the system with non-zero charge
rotates, a magnetic field might be created at relatively later stage.
Such late-stage magnetic fields may also contribute to the observed
polarization~\cite{Guo:2019mgh}.

In the global polarization picture based on the vorticity, one expects
different particles to be polarized depending only on the particle
spin in accordance with Eq.~\eqref{eq:polar}.  A deviation could arise
from effects of the initial magnetic field mentioned above, and from
the fact that different particles are produced at different times or
regions as the system freezes out~\cite{Vitiuk:2019rfv}, or through
meson-baryon interactions~\cite{Csernai:2018yok}.  Therefore, to
understand the nature of the global polarization, it is crucial to
measure the polarization of different particles, and if possible,
particles with different spins.  The polarization measurement with
particles of different magnetic moments would provide additional
information on the magnitude of the magnetic field.  For example, the
magnetic moment of $\Omega$ hyperon is three times larger than that of
$\Lambda$ hyperon ($\mu_{\Omega^-}=-2.02\mu_N$ and
$\mu_{\Lambda}=-0.613\mu_N$ where $\mu_N$ is the nuclear magneton).
Thus $\Omega$ hyperons are more sensitive to the magnetic field
contribution to the polarization.

Comparing the polarization between particles and antiparticles, it
might be important to account for the effects of non-zero baryon
chemical potential~\cite{Fang:2016vpj}, especially at lower collisions
energies.  Besides directly affecting the quark distributions, the
non-zero chemical potential could lead to different freeze-out
conditions, and thus to different effective vorticities responsible
for the particle polarization. But overall such effects are expected
to be relatively small.

%------------------------------------------------------------------
\subsection{Anisotropic flow and polarization along the beam
  direction%, $\pz$.
}
\label{sec:ani-flow}

Anisotropic flow leads to non-trivial collective velocity fields in
the transverse direction, which in its turn would manifest itself via
particle polarization along the beam
direction~\cite{Voloshin:2017kqp}. The polarization $\pz$ component
dependence on the azimuthal angle would in general follow the
anisotropic flow pattern of the same harmonic. We use a simple Blast
Wave model to illustrate this phenomenon below.

\begin{figure}[th]
\centerline{
	\hspace{0.02\linewidth}
	\begin{minipage}[b]{0.46\linewidth}
	\includegraphics[width=\columnwidth]{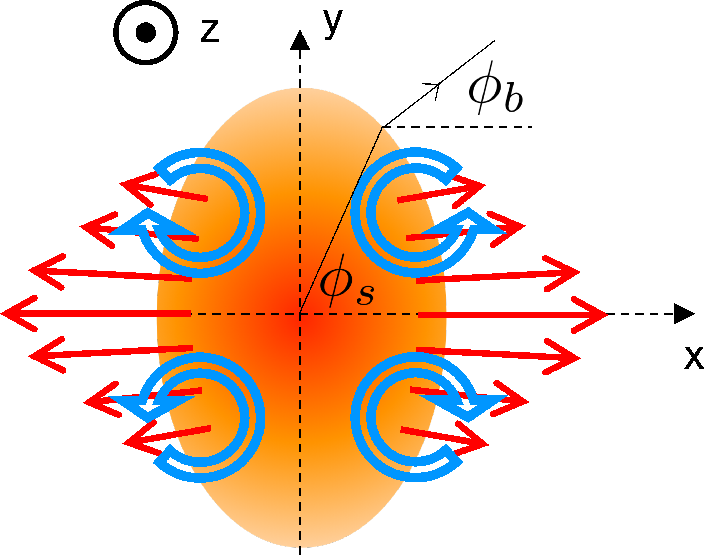}
	\vspace{+1pt}
	\end{minipage}
	\hspace{0.04\linewidth}
	\begin{minipage}[b]{0.47\linewidth}
	\includegraphics[width=\columnwidth]{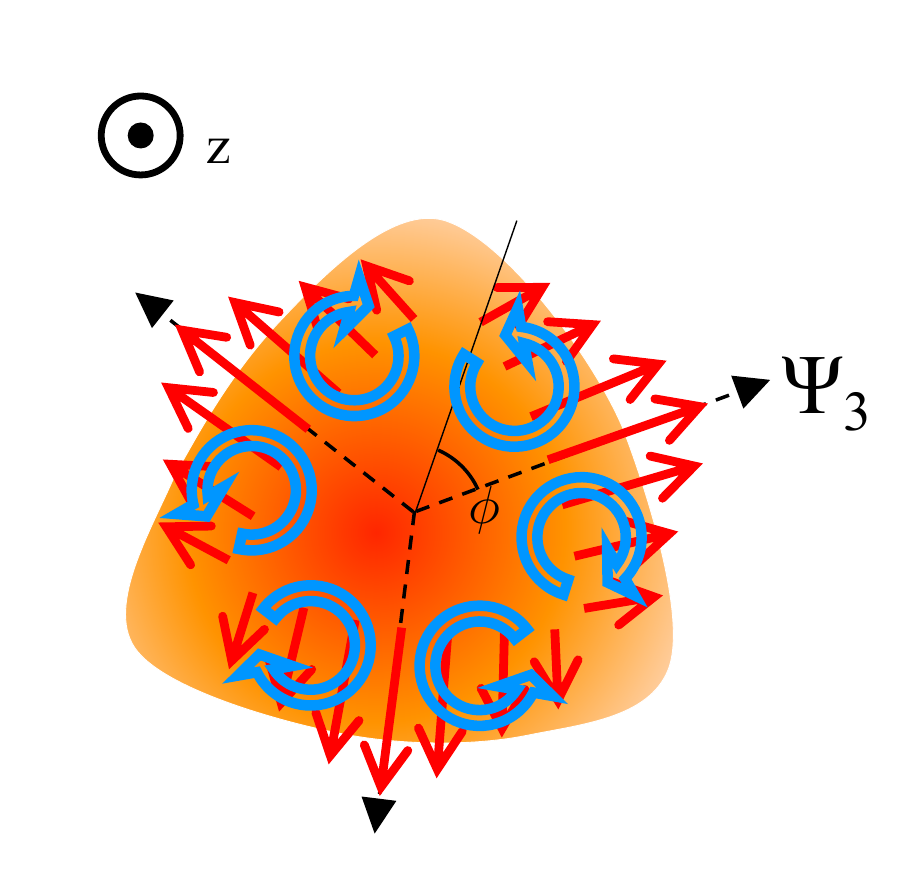}
	\end{minipage}
}
\caption{ Transverse plane schematic view of the system leading to
  elliptic (left) and triangular flow (right). Red solid arrows indicate the
  expansion velocity, the largest along the greatest density gradients
  defining the flow angles, $\Psi_2$ (coincides with $x$-axis) and
  $\Psi_3$. In the left sketch, $\phi_s$ denotes azimuthal angle of
  the source element which is boosted to $\phi_b$ direction. 
  Blue open arrows indicate local vorticities induced by the anisotropic flow. }
\label{fig:BW23}
\end{figure}

In a simplest version of the Blast-Wave model including
anisotropic flow~\cite{STAR:2001ksn,Voloshin:2011mg,Voloshin:2017kqp},
the particle production source at freeze-out is parameterized with 5
parameters: temperature $T$, maximum transverse radial flow velocity
(rapidity) $\rho_{t,max}$ and amplitude of the azimuthal modulation in
expansion velocity denoted as $b_n$, parameter $R$ characterizing the
size, and the spatial anisotropy parameter $a_n$. The source, see
Fig.~\ref{fig:BW23}(left), is then described by the following equations:
%\red{TN: I guess $\cos(n\phi_s)$ in $\rho_t$ should be $\cos(n\phi_b)$?}
%
\be
r_{max} = R[1-a_n \cos(n\phi_s)],
\hspace{0.5 cm}
\rho_t = \rho_{t,max}\left[\frac{r}{r_{max}(\phi_s)}\right][1+b_n\cos(n\phi_s)].
\label{eq:boost}
\ee
It is assumed that the source element located at azimuthal angle
$\phi_s$ is boosted with velocity $\rho_t$ perpendicular to the
surface of the source ($\phi_b$). Note that the parameters $a_n$ and
$b_n$ are usually small, $a_n,\, b_n < 0.1$, as follows from comparison
of the model to the anisotropic flow and azimuthally sensitive
femtoscopic measurements~\cite{STAR:2004qya}. In the limit
of  $a_n\ll1,\ b_n\ll 1$ , the longitudinal component of the vorticity
can be written as:
\be
\omega_z = \frac{1}{2} (\grad \times
\bv)_z \approx \left(\frac{\rho_{t,nmax}}{R}\right)\sin(n\phi_s)[b_n-a_n].
\label{eq:vortn}
\ee
It results in the following estimate for the hyperon polarization:
\be
P_z \approx \frac{\omega_z}{2T}\approx 0.1\sin(n\phi_s)[b_n-a_n],
\ee
where it is assumed that $\rho_{t,nmax}\sim1$, $R\sim 10$~fm, and
$T\sim 100$~MeV.  In practice, the coefficients $b_n$ and $a_n$ are
both of the order of a few percent~\cite{STAR:2004qya}, often close to
each other. That results in the magnitudes of $z$-polarization not
greater than a few per-mill, almost an order of magnitude lower than
what was obtained in original hydrodynamics
calculations~\cite{Becattini:2017gcx}.
Note that the estimate above is valid for anisotropic flow of any
harmonics $n$.

%----------------------------
\subsection{Circular polarization, $P_\phi$;
  polarization along $x$-direction, $\px$}
\label{sec:circular}

\begin{figure}[th]
\centerline{
	\includegraphics[width=0.6\columnwidth]{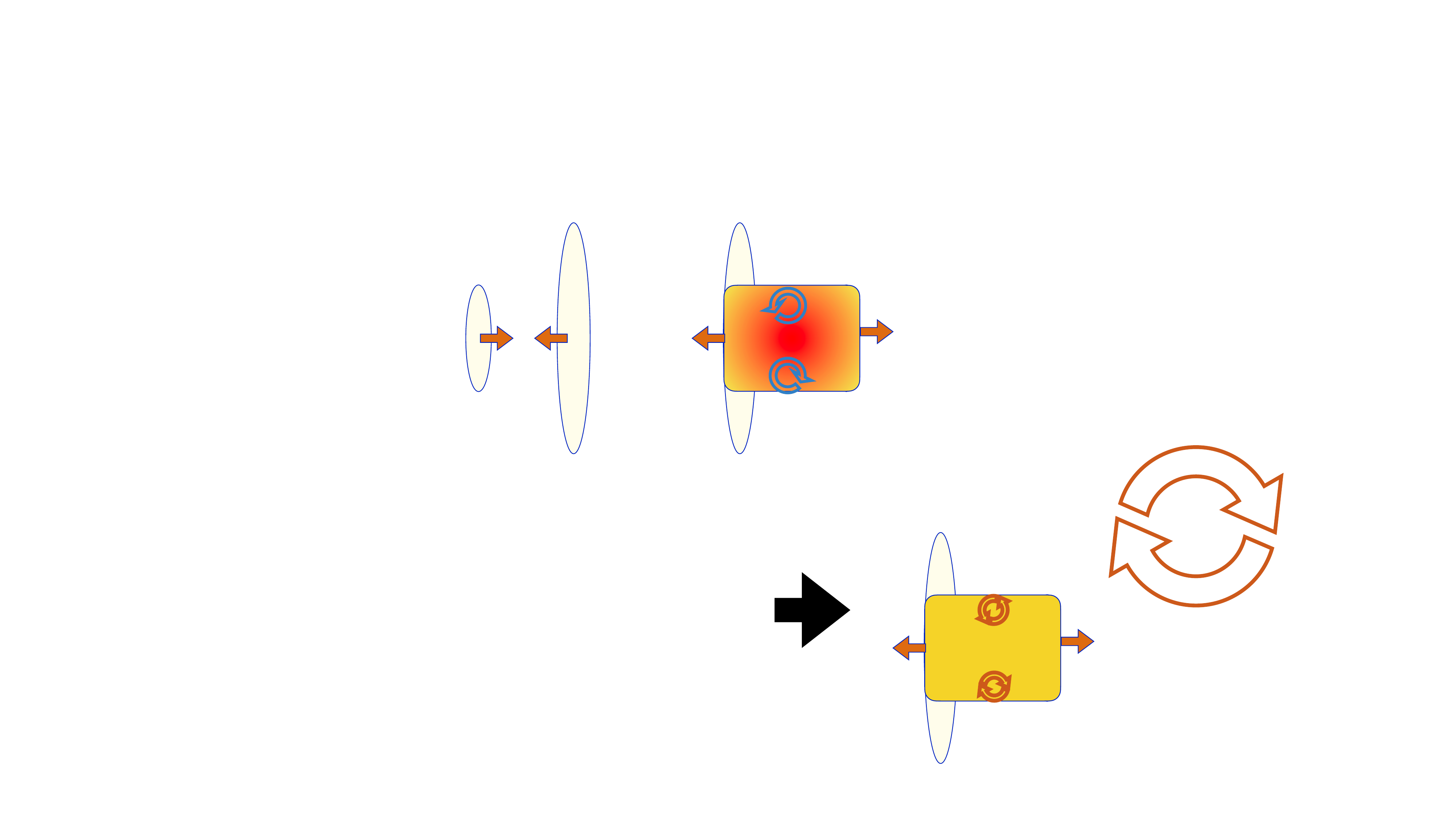}
}
\caption{ A schematic view of a central asymmetric collision, such as
  Cu+Au, before (left) and after (right) the collision. The blue open 
  arrows indicate the vorticity at the outer edges of the collision zone.  }
\label{fig:asymmetric}
\end{figure}

Non-uniform stopping in the transverse plane, and dependence of the
expansion velocity on rapidity leads to toroidal structure of the
velocity field~\cite{Xia:2018tes,Ivanov:2018eej}. Theoretical
calculations suggest that a vortex ring could be created at very
forward/backward regions, most prominent in the central collisions.
Even a better pronounced vortex ring could be created when smaller
object passes through larger object such as central asymmetric
collisions of Cu+Au, $d$+Au, and $p$+Au as first proposed in
Ref.~\cite{Voloshin:2017kqp}, see Fig.~\ref{fig:asymmetric}.  The
calculations~\cite{Lisa:2021zkj} show that the polarization due to
such vortex rings could reach values as high as a few percent.  The
smaller object could be replaced with a jet instead of
nuclei~\cite{Serenone:2021zef}. According to the simulation of jet
interacting with medium~\cite{Betz:2007kg,Tachibana:2012sa}, vortex
rings can be created around the path which jet passes through in the
medium.

The axis of such a vortex ring is along the azimuthal direction
relative to the smaller-nucleus-going or jet-going direction.  The
expected polarization in case of central A+B collisions can be
expressed as $P_{\phi}\propto \hat{p}_{\rm T} \times \hat{z}$ where
$\hat{p}_T$ and $\hat{z}$ are the unit vectors along the particle transverse
momentum and the beam direction, respectively (replace $\hat{z}$ with
a unit vector pointing along the jet axis in case for the jet-induced
polarization).  Such measurement would require a careful treatment of the
detector acceptance effects, excluding left-right asymmetry in
particle reconstruction.

%-------------------
%\subsection{x-component of polarization: $P_{\rm x}$}

Finally we note the importance of the polarization measurements along
the impact parameter direction, $P_{\rm x}$.  Theoretical
models~\cite{Xia:2018tes,Karpenko:2016jyx} suggest that $P_{\rm x}$
has an azimuthal dependence following $\sin(2\phi)$ curve where $\phi$
is the hyperon's azimuthal angle relative to the reaction plane.  Such
a dependence on the azimuthal angle is also expected from simple
calculations based on the Glauber model, see right panel in
Fig.~\ref{fig:glauber} where $\omega_x\propto \frac{1}{2}\partial
v_z/\partial y$.  This component could also have a contribution from
the so-called shear induced polarization, see the discussion in
Sec.~\ref{sec:SIP-SHE}.
%\red{do you want to mention something expected from SIP?}
Similarly to the $P_\phi$ measurements, these measurements could be
technically difficult accounting for the acceptance effect.

%====================================
%\section{Modeling} \label{sec:modeling}
%==================================================================
%\section{Modeling}
\section{Spin and polarization in hydrodynamic description}

%----------------------------------------------------------
%\subsection{Spin and polarization in hydrodynamic description}
\subsection{Kinematic vorticity, thermal gradients, acceleration}

The Blast Wave model described in Sec.~\ref{sec:ani-flow} is likely a
gross oversimplification of the reality.  It accounts, though still
approximately, only for the contribution form so-called kinematic
vorticity neglecting several other potentially important
contributions. At the same time, as we discuss below,
it describes surprisingly well the main features of the data.
In relativistic hydrodynamics, the mean spin vector of
$s=1/2$ particles with mass $m$ and four-momentum $p$ is given by the
following equation~\cite{Becattini:2013fla}:
\begin{equation}\label{basic}
  S^\mu(x,p)= - \frac{1}{8m} (1-n_F) 
\epsilon^{\mu\tau\rho\sigma} p_\tau \varpi_{\rho\sigma},
\end{equation}
where $n_F$ is the Fermi-Dirac distribution and $\tomega$ is the {\it
  thermal vorticity} defined as:
\begin{equation}\label{thvort}
  \tomega_{\mu\nu} = 
\frac{1}{2} \left[ \pd_\nu (u_\mu/T) - \pd_\mu (u_\nu/T) \right];
\end{equation}
 $u_\mu =(\gamma,\gamma\,\bv)$ is the fluid 4-velocity ($\gamma$ is
the gamma factor), and $T$ is the proper temperature.  Thermal
vorticity can be subdivided into two parts - the part including the
temperature gradients $\pd_\mu(1/T)$ and the part proportional to the
kinematic vorticity $ \omega_{\mu\nu}^{\rm K} = \frac{1}{2} \left(
\pd_\nu u_\mu - \pd_\mu u_\nu \right)$.  The latter in its turn can be
separated into the part proportional to the acceleration and the
spatial (transverse) part:
\be
\omega_{\mu\nu}^{\rm K}  =
\frac{1}{2} \left( A_\mu u_\nu - A_\nu u_\mu \right) +
\frac{1}{2} \left( \nabla_\nu u_\mu - \nabla_\mu u_\nu \right),
\ee
where the acceleration vector $A^\mu = D u^\mu$ ($D=u^\nu\pd_\nu$ is
the co-moving time derivative) and the $\nabla_\mu=\pd_\mu -u_\mu D$
is the so-called ``orthogonal'' (to $u^\mu$) derivative.  The
transverse part of the kinematic vorticity can be also expressed via
the vorticity (angular velocity) vector
\be \omega^\mu = \frac{1}{2}
\epsilon^{\mu\nu\rho\sigma} u_\nu \pd_\rho u_\sigma,
\ee 
as 
\be
\frac{1}{2} \left( \nabla_\nu u_\mu - \nabla_\mu u_\nu \right) =
\epsilon_{\mu\nu\rho\sigma} \omega^\rho u^\sigma. 
\ee
Using these notations and combining everything together, 
the spin vector can be written as 
\begin{eqnarray}\label{spindeco}
&&S^\mu(x,p) =
 \frac{1}{8m} (1-n_F) 
\nonumber
\\ &&\times \left[ 
- \epsilon^{\mu\nu\rho\sigma}
p_\sigma \frac{1}{T^2} (\pd_\nu T) u_\rho  +  2 \,
\frac{\omega^\mu (u^\nu p_\nu) - u^\mu (\omega^\nu  p_\nu)}{T} -
\frac{1}{T} \epsilon^{\mu\nu\rho\sigma} p_\sigma
A_\nu u_\rho \right],
\end{eqnarray}
with three terms describing contributions of the temperature gradient,
the vorticity, and the acceleration, respectively.  Note that in an
ideal uncharged fluid the temperature gradient term and the
acceleration contribution are related by the equation of motion
$\nabla_\mu T = T A_\mu$.

It is instructive to rewrite the expression Eq.~\ref{spindeco} in the
rest frame of the fluid, where $u^\mu=(1,0,0,0)$, $D=\pd^t$,
$\nabla_\mu=(0,\grad)$, and $\omega^\mu=(0,\bomega)$:
%Also, for simplicity we assume $n_F\ll 1$ below.
%
\begin{eqnarray}\label{spindeco}
&&S^0(x,p) =  \frac{1}{8m} (1-n_F) \, \frac{\bomega\cdot\bp}{T}, 
\\ && \bS(x,p) =\frac{1}{8m} (1-n_F)
\left( 
- \frac{\bp \times \grad T}{T^2}  
+  2 \, \frac{E \, \bomega }{T} 
+\frac{\bp \times \bA}{T} 
\right).
\end{eqnarray}
In the above expressions, $E$ and $\bp$ are the energy and momentum of
the particle in the fluid rest frame.  In the nonrelativistic limit,
the contribution related to the angular velocity (coinciding with
nonrelativistic vorticity) is the largest, with the contribution from
temperature gradients and acceleration being suppressed by $v/c$
powers.

For completeness we also present an equation for the average spin
vector transformation
\be
\bS^*=\bS-\frac{\bp \cdot \bS}{E(E+m)}\bp,
\ee
which should be used for calculation of the spin vector in the
particle rest frame.

%-----------------------------------------------------------
\subsection{Shear-induced polarization and the spin Hall effect}
\label{sec:SIP-SHE}

Very recently two groups~\cite{Liu:2021uhn,Becattini:2021suc}
independently reported a new mechanism for the spin polarization - so
called ``shear induced polarization'' (SIP) originated in symmetric
part of the velocity gradients $\ximunu=1/2 (\dmu \unu +\dnu \umu)$.
Note that the expression for the polarization due to symmetric part of
the velocity gradient tensor obtained by two groups are similar but
not exactly the same with one qualitative difference as that the
expressions obtained in Ref.~\cite{Becattini:2021suc} explicitly
depends on the freeze-out hyper-surface shape, while the expression in
Ref.~\cite{Liu:2021uhn} allows ``local'' interpretation. For our
qualitative discussion of the effect below we will use the definition
in Ref.~\cite{Liu:2021uhn}.

The origin of SIP is the motion of a particle in anisotropic fluid. It
is zero if the particle is moving with the fluid velocity, which is in
contrast to the polarization due to vorticity. It is clearly seen if
the corresponding expressions are written in the fluid rest frame
$\umu=(1,0,0,0)$:
\begin{equation}
S_i^{({\rm vort)}} \approx \frac{E}{8mT}\; \epsilon_{ikj}\frac{1}{2}(\pd_k
v_j-\pd_j v_k),
\end{equation}
\begin{equation}
S_i^{({\rm shear})} \approx \frac{1}{4mTE} \; \epsilon_{ikj}
p_k p_m \, \frac{1}{2}(\pd_j v_m+\pd_m v_j).
\end{equation}
One can see that the SIP contribution to the polarization is
suppressed by the order of $(p/E)^2$ compared to vorticity
contribution and become zero for particles moving with the fluid
velocity.  It was also pointed out recently that the chemical
potential gradients could also contribute to the polarization.  This
contribution identified as the ``spin Hall effect''
(SHE)~\cite{Liu:2020dxg}. In the fluid rest frame :
\begin{equation}
S_i^{({\rm SHE})} \approx \frac{1}{4mE} \; \epsilon_{ikj}
p_k  \, \pd_j (\mu/T).
\end{equation}
Similarly to SIP, the polarization due to the gradients in baryon
chemical potential (SHE -- spin Hall effect) is also suppressed by a
power of $p/E$.  The SHE might be important in particular for the
interpretation of the difference in polarization of particles and
antiparticles. Note that the role of chemical potential was also
studied earlier in a different content with the conclusion that for
nonrelativistic hyperons the effect is almost
negligible~\cite{Fang:2016vpj}. The new effects, both SIP and SHE, are
related to the motion of the particle in anisotropic fluid, and thus
expected to be small for nonrelativistic particles compared to the
polarization due to vorticity. This is the main reason why the
calculations that involve quark degree of freedom result in stronger
effects compared to those where SIP and SHE contributions are
calculated directly for (nonrelativistic) hyperons. We discuss this in
more detail in relation to the experimental measurements of the
polarization along the beam direction ($\pz$).

%------------------------------------------------
\subsection{Additional comments}

While several model calculations do show a significant contribution to
the hyperon polarization from temperature gradients and acceleration,
in our more qualitative discussion we mostly argue on the basis of the
contribution from kinematic vorticity.
%Almost all of our discussion is about hyperon polarization.
The freeze-out temperature of the system is about $\sim 100$~MeV, and
all the hyperons are nonrelativistic in the local fluid frame.  For
that reason we also often estimate the polarization in the fluid
frame, although all the experimental measurements are performed in the
particle rest frame. We do treat the hyperons as relativistic in the
laboratory frame though, as the fluid collective motion is
relativistic. The nonrelativistic treatment might fail if the final
particle polarization is due to the coalescence of initially (during
the system evolution before the hadronization) polarized (constituent)
quarks, with masses that are only factor of 2--3 higher than the
temperature.

All hydrodynamic calculations use the Cooper-Frye
prescription~\cite{Cooper:1974mv} for the fluid freeze-out.  This
prescription has several known problems (see e.g.,
Refs.~\cite{Rischke:1998fq,Magas:1999zm}), which might be not very
important for calculations of the particle spectra, but it is not
known how good the Cooper-Frye prescription is for calculation of the
polarization. In particular, the contributions from the temperature
gradients and acceleration might be questionable, as the very concept
of freeze-out assumes insignificance of those effect. Then. their
contributions would be related to the corresponding relaxation times
of the system.  Note, that if found significant, the measurement of
those effects might provide unique information about the velocity and
temperature gradients at freeze-out, for which the particle spectra
are mostly insensitive.

Vortical effects may also strongly affect the baryon dynamics of the
system, leading to a separation of baryon and antibaryons along the
vorticity direction (perpendicular to the reaction plane) -- the
so-called Chiral Vortical Effect (CVE)~\cite{Kharzeev:2015znc}. The
CVE is similar in many respect to the more familiar Chiral Magnetic
Effect (CME) - the electric charge separation along the magnetic
field. For reviews on those and similar effects, as well as the status
of the experimental search for those phenomena, see
Refs.~\cite{Kharzeev:2015znc,Koch:2016pzl}. For a reliable theoretical
calculation of both effects, one has to know the vorticity of the
created system as well as the evolution of (electro)magnetic field.

In view of the recent polarization measurements in ultra-relativistic
heavy-ion collisions, note the discussion~\cite{Becattini:2020riu}
of a physical meaning of the spin angular momentum in quantum field theory
and relativistic hydrodynamics.

%=====================
\section{How is it measured} \label{sec:measurements}
%==============================================================
%\section{How is it measured}

%-----------------------------------------------------------
\subsection{Self-analyzing weak decays of hyperons}\label{sec:self-ana}

The hyperon weak decays provide a most straightforward way to
experimentally measure polarization of particles produced in heavy-ion
collisions. Because of its parity-violating weak decay, the angular
distribution of the decay products at the hyperon rest frame obeys the
following relation:
\be
  \frac{dN}{d\Omega^{\ast}} =\frac{1}{4\pi} \left(1 
+ \alpha_{\rm H} {\bm P}_{\rm H}^{\ast}\cdot \hat{\bm p}_{\rm B}^\ast \right), 
\label{eq:hyperon_decay}
\ee
where $\alpha_{\rm H}$ is the hyperon decay parameter, ${\bm P}_{\rm
  H}^*$ is the hyperon polarization vector, and $\hat{\bm p}_{\rm
  B}^*$ is the unit vector in the direction of the daughter baryon
momentum, and the $\Omega$ is its solid angle.  The asterisk is used
to denote quantities in the hyperon rest frame.  The decay parameter
$\alpha_{\rm H}$ reflects analyzing power in the measurement, it is
different for different hyperons.  In case of \lam hyperon decay of
$\lam\rightarrow p+\pi^-$, the decay parameter is
$\alpha_{\lam}=0.732\pm0.014$~\cite{ParticleDataGroup:2022pth}.  The
decay parameter for $\alam\rightarrow \bar{p}+\pi^+$ is usually
treated same as \lam, i.e., $\alpha_{\lam}=-\alpha_{\alam}$, assuming
that charge conjugation parity (CP) symmetry is conserved, although
the world average data shows slightly higher value
$\alpha_{\alam}=-0.758\pm0.014$~\cite{ParticleDataGroup:2022pth}.  In
this paper, we follow this convention unless it is specified in the
figure.  We also note that the recent studies
~\cite{BESIII:2021ypr,BESIII:2022qax} indicate that $\alpha_{\lam}$
could be be larger by a few percent compared to the aforementioned
value (and be closer to $\alpha_{\lam}\approx\alpha_{\alam}$), leading
to a few \% reduction of the measured polarization.

%--------------------------------------------------------------------
\subsubsection{Multistrange hyperons and two-step decays}
\label{sec:multistrange}

Multistrange hyperons such as $\Xi$ and $\Omega$ decay in two steps.
For example, in case of $\xim$ hyperon (spin-$1/2$), $\xim \rightarrow
\lam +\pi^-$ with subsequent decay $\lam \rightarrow p+ \pi^-$. If
$\xim$ is polarized, its polarization is partially transferred to the
daughter $\lam$.  Both steps in such a cascade decay are
parity-violating and thus can be used for an independent measurement
of the parent hyperon polarization. The decay parameter for $\xim
\rightarrow \Lambda +\pi^-$ is $\alpha_{\xim}=-0.401\pm
0.010$~\cite{ParticleDataGroup:2022pth}.  This value of $\alpha_{\Xi}$
was constrained by the measurement of the product of
$\alpha_{\Xi}\,\alpha_{\lam}$ with the $\alpha_{\lam}$ measured
separately, therefore the change of $\alpha_{\lam}$ would affect the
value of $\alpha_{\Xi}$ as well.  We also note that the recent direct
measurement of $\alpha_{\Xi}$~\cite{BESIII:2021ypr} suggests a
slightly different value ($\alpha_{\Xi}=-0.376\pm0.007$).

The polarization of the daughter baryon in the weak decay of a
spin-$1/2$ hyperon is described by the Lee-Yang
formula~\cite{Lee:1957qs,Luk:2000zw,Huang:2004jp} with three decay
parameters; $\alpha$, $\beta$, and $\gamma$, where $\alpha$ is a
parity-violating part reflecting decay asymmetry as mentioned above,
$\beta$ accounts for the violation of the time reversal symmetry, and
$\gamma$ satisfies $\alpha^2+\beta^2+\gamma^2=1$.  For a particular
case of $\Xi\rightarrow \Lambda + \pi$ decay, the daughter \lam
polarization in its rest frame can be written as:
\begin{equation}\label{eq:poldecay}
{\bf P}^*_{\Lambda}=
\frac{(\alpha_{\Xi}+{\bf P}^*_{\Xi}
  \cdot\hat{\bm p}_\Lambda^*)\hat{\bm p}_\Lambda^* 
  + \beta_{\Xi}{\bf P}^*_{\Xi}\times\hat{\bm p}_\Lambda^* 
    + \gamma_{\Xi} \hat{\bm p}_\Lambda^*\times({\bf P}^*_{\Xi}\times\hat{\bm
    p}_\Lambda^*)}{1+\alpha_{\Xi}{\bf P}^*_{\Xi} \cdot\hat{\bm p}_\Lambda^*} ,
\end{equation}
where $\hat{\bm p}_\Lambda^*$ is the unit vector of the $\Lambda$
momentum, and $\Xi$ polarization ${\bm P}^*_{\Xi}$ is given in the
$\Xi$ rest frame. Averaging over the angular distribution of the
daughter $\Lambda$ in the rest frame of the $\Xi$ given by
Eq.~\eqref{eq:hyperon_decay} leads to
\be
{\bf  P}^*_{\Lambda} = C_{\Xi^- \Lambda} {\bf P}^*_{\Xi} =
\tfrac{1}{3} \left( 1+2\gamma_{\Xi} \right) {\bf P}^*_{\Xi}.  \label{eq:PlamPxi}
\ee
Using the measured value for the $\gamma_{\Xi}$
parameter~\cite{ParticleDataGroup:2022pth,Huang:2004jp}, the
polarization transfer coefficient for $\Xi^{-}$ to $\Lambda$ decay
leads to:
\begin{align}
C_{\Xi^- \Lambda}=&\tfrac{1}{3}\left(1+2\times0.916\right)=+0.944. 
%\nonumber \\ 
%&\tfrac{1}{3}\left(2\times0.85+1\right)=+0.900
\end{align}
This shows that the polarization of $\Xi^-$ is transferred to daughter
$\Lambda$ almost at its full value.  We also would like to point out
that the value of the $\gamma$ parameter is constrained by the
measured $\alpha$ and $\phi$ as $\gamma=(1-\alpha^2)\cos^2\phi$ where
$\phi=\tan^{-1}\beta/\gamma$, therefore the change in $\alpha$
parameter as well as $\phi$ value would also lead to a change in the
$\gamma$ parameter.

The polarization of the daughter baryon in a two-particle decay of
spin-$3/2$ hyperon, i.e., $\Omega\rightarrow \Lambda +K$, can be also
described by three parameters $\alpha_\Omega$, $\beta_\Omega$, and
$\gamma_\Omega$~\cite{Luk:1988as}.  The decay parameter
$\alpha_\Omega$ determines the angular distribution of \lam in the
$\Omega$ rest frame and is measured to be
small~\cite{ParticleDataGroup:2022pth}:
$\alpha_\Omega=0.0157\pm0.0021$; this means that the measurement of
$\Omega$ polarization via analysis of the daughter $\Lambda$ angular
distribution is practically impossible.  The polarization transfer in
this decay is determined by the $\gamma_\Omega$ parameter
as~\cite{Luk:1983pe,Luk:1988as,Kim:1992az}:
\be 
{\bf  P}^*_{\Lambda} = C_{\Omega^- \Lambda} {\bf P}^*_{\Omega} =
\tfrac{1}{5} \left( 1+4\gamma_{\Omega} \right) {\bf P}^*_{\Omega}.  
\label{eq:PlamPom}
\ee
The parameter $\gamma_\Omega$ is unknown but considering the
time-reversal violation parameter $\beta_\Omega $ to be small, one can
expect that the unmeasured parameter $\gamma_\Omega$ is $\gamma_\Omega
\approx \pm 1$, based on the constraint
$\alpha^2+\beta^2+\gamma^2=1$. This results in a polarization transfer
to be $C_{\Omega^- \Lambda} \approx 1$ or $C_{\Omega^- \Lambda}\approx
-0.6$. As discussed later, the measurement of $\Omega$ global
polarization can help to resolve the ambiguity of $\gamma_\Omega$
under assumption of the vorticity picture in heavy-ion collisions.

%-----------------------------------------------------------------
\subsection{Global polarization measurement}

Polarization component along the initial orbital angular momentum
${\bm L}$ for hyperons, referred to as global polarization when
averaged over all produced particles, can be obtained by integrating
Eq.~\eqref{eq:hyperon_decay} over the polar angle of daughter baryon
$\theta_{\rm B}^{\ast}$ and the reaction plane angle $\Psi_{\rm RP}$,
considering the projection of the polarization onto the direction
${\bm L}$.  In Eq.~\eqref{eq:hyperon_decay}, ${\bm P}_{\rm
  H}^{\ast}\cdot \hat{\bm p}_{\rm B}^\ast$ can be substituted with
$P_{\rm H}\cos\theta^\ast=P_{\rm H}\sin\theta_{\rm
  B}^\ast\sin(\Psi_{\rm RP}-\phi_{\rm B}^\ast)$ where $\theta^\ast$ is
the angle between the polarization vector and momentum of daughter
baryon in the hyperon rest frame, and $\theta_{\rm B}^\ast$ and
$\phi_{\rm B}^\ast$ are azimuthal and polar angles of daughter baryon
in the hyperon rest frame.  Then the average of $\sin(\Psi_{\rm
  RP}-\phi_{\rm B}^\ast)$ is calculated as
\begin{eqnarray}
  & &\hspace{-1em}
  \langle\sin(\Psi_{\rm RP}-\phi_{\rm B}^\ast)\rangle \nonumber\\
  &=& \int d\Omega^\ast \int \frac{d\Psi_{\rm RP}}{2\pi} 
  \frac{1}{4\pi}
  \left[ 1+\alpha_{\rm H}P_{\rm H}\sin\theta^\ast_{\rm
      B}\sin(\Psi_{\rm RP}-\phi_{B}^\ast) \right]
  \sin(\Psi_{\rm RP}-\phi_{B}^\ast) \\
  &=& \frac{\alpha_{\rm H}P_{\rm H}}{2} \int d\Omega^\ast \sin\theta^\ast_{\rm B}.
\end{eqnarray}
Hence, this leads to
\begin{eqnarray}
P_{\rm H} 
&=& \frac{8}{\pi\alpha_{\rm H}}
\frac{1}{(4/\pi)
  \int d\Omega^\ast \sin\theta_{\rm B}^\ast} \langle\sin(\Psi_{\rm
  RP}-\phi_{\rm B}^\ast)\rangle
\nonumber
\\
&=& \frac{8}{\pi\alpha_{\rm H}} \frac{1}{A_{0}} \langle\sin(\Psi_{\rm
  RP}-\phi_{\rm B}^\ast)\rangle,
\label{eq:PH_nores}
\end{eqnarray}
where $A_0=(4/\pi) \int d\Omega^\ast \sin\theta_{\rm B}^\ast$ is an
acceptance correction factor and usually estimated in a data-driven
way. If the detector can measure all produced hyperons of interest,
the factor $A_{0}$ becomes unity. In practice, $A_0$ slightly deviates
from unity and depends on the event multiplicity and the hyperon
transverse momentum $p_T$.  Equation~\eqref{eq:PH_nores} does not
account for the polarization dependence on azimuthal angle, but such a
dependence as well as the presence of elliptic flow might affect the
measurements.  For a more detailed discussion of the acceptance and
tracking efficiency corrections including $P_{\rm H}$ azimuthal
dependence, see Sec.~\ref{sec:DetAcc}.

Experimentally the first-order event plane angle $\Psi_{1}$ is used as
a proxy of $\Psi_{\rm RP}$.  Then Eq.~\eqref{eq:PH_nores} can be
rewritten to take into account for the event plane resolution as
follows~\cite{STAR:2007ccu}
\begin{eqnarray}
  P_{\rm H} = \frac{8}{\pi\alpha_{\rm H}}
  \frac{1}{A_0}
  \frac{\langle\sin(\Psi_1-\phi_{\rm B}^\ast)\rangle}{\rm Res(\Psi_1)}, 
\label{eq:PH_res}
\end{eqnarray}
where ${\rm Res(\Psi_1)}$ is the event plane resolution defined as
$\langle\cos(\Psi_1-\Psi_{\rm RP})\rangle$.  The azimuthal angle
$\Psi_1$ can be determined by measuring spectator fragments and
provides the direction of the initial orbital angular
momentum~\cite{Voloshin:2016ppr}.

Equation~\ref{eq:PH_res} provides a possibility to measure global
polarization of hyperons by measuring only azimuthal distributions of the
daughter baryon. While this approach based on well established
anisotropic flow techniques, a slightly better statistical accuracy
could be achieved by measuring the full angular distribution,
including polar angle. In this case,
\be
P_{\rm H} = \frac{3}{\alpha_{\rm H}} \mean {\cos\theta^\ast} =
  \frac{3}{\alpha_{\rm H}}
  \mean{ \sin\theta_{\rm  B}^\ast
    \sin(\Psi_{\rm RP}-\phi_{\rm B}^\ast )}.
  \label{eq:poltheta}
\ee
For a  discussion of the acceptance
and tracking efficiency corrections including the azimuthal
dependence, see Sec.~\ref{sec:DetAcc}.

%--------------------------------------------------------------------
\subsubsection{Global polarization in which frame?}

As defined in Eq.~\eqref{eq:hyperon_decay}, the polarization is
measured in the hyperon rest frame and the global polarization is the
polarization component along the orbital angular momentum ${\textbf
  {\it L}}$ direction in the center-of-mass frame of heavy-ion
collisions as shown in Eq.~\eqref{eq:PH_res}.  Strictly speaking, the
${\textbf {\it L}}$ direction in $\Lambda$ rest frame is different
from the ${\textbf {\it L}}$ direction in the center-of-mass frame of
heavy-ion collisions, and therefore the proper treatment of the
reference frame and measured polarization is needed.
Reference~\cite{Florkowski:2021pkp} studies the effect of the frame
difference in the measurement of global polarization and found that
the effect is small and reaches about 10\% for high transverse
momentum ($\pt\sim$ 4--5 GeV/$c$).  Note that the mean $p_T$ for
$\Lambda$ is $\sim1$ GeV/$c$~\cite{Adam:2019koz} depending on the
centrality and collision energy, and could be slightly higher
($\pt\sim1.5$~GeV/$c$) with the kinematic cut for $\Lambda$ used in
the polarization measurement.

%--------------------------------------------------------------------
\subsection{Measuring polarization induced by anisotropic flow}

As already mentioned in Sec.~\ref{sec:ani-flow}, one can expect that
azimuthal anisotropic flow would lead to a vorticity pointing along
the beam direction. The orientation, along or opposite to the beam, in
this case depends on the azimuthal angle of the particle.  Similarly
to the case of the global polarization, the longitudinal component of
the polarization $\pz$ can be obtained by integrating
Eq.~\eqref{eq:hyperon_decay} as $\int
dN/d\Omega^{\ast}\cos\theta^{\ast} d\Omega^{\ast}$ in which ${\bm
  P}_{\rm H}\cdot \hat{p}_{\rm B}^\ast$ is replaced by $P_{\rm
  z}\cos\theta^{\ast}_{\rm B}$, where $\theta^{\ast}_{\rm B}$ is the
polar angle of daughter baryon in the parent hyperon rest frame. This
leads to
\begin{eqnarray}
  P_{\rm z} = \frac{1}{A_{\rm z}}
  \frac{3 \langle\cos\theta_{\rm B}^\ast\rangle}{\alpha_{\rm H}},
\label{eq:pz}
\end{eqnarray}
where $A_{\rm z}$ is the acceptance correction factor (see
Sec.~\ref{sec:acc-Pz}).  For the case of perfect detector acceptance,
$A_{\rm z}=1$.  The factor $A_{\rm z}$ can be calculated using the
data and is known to weekly depend on the transverse momentum and
collision centrality. For more details on the correction factors,
see Sec.~\ref{sec:DetAcc}.
%Note that the correction factor also depends on
%the signal itself as seen in Eq.~\eqref{eq:Az} but for small signal it
%will be a constant value close to unity to a first approximation.

As follows from considerations in Sec.~\ref{sec:ani-flow}, see
Eq.~\ref{eq:vortn}, the polarization induced by $n$-th harmonic
anisotropic flow, if any, is expected to depend on the azimuthal angle
of hyperons as $P_z(\phi)\propto\sin[n(\phi-\Psi_n)]$. Then such a
polarization can be quantified by the corresponding Fourier
coefficient
\begin{equation}
  P_{z,sn}=\mean{P_z\,\sin[n(\phi_H-\Psi_{n})]},
\label{eq:Pzsin}
\end{equation}
where $\Psi_n$ is n-th harmonic event plane. Then, similarly to
the case for global polarization as in Eq.~\eqref{eq:PH_res}, 
Eq.~\eqref{eq:Pzsin} can be rewritten accounting for the event plane resolution
\begin{equation}
  P_{z,sn}=\frac{\mean{P_z\,\sin[n(\phi_H-\Psi_{n})]}}{\rm Res(n\Psi_n)},
\label{eq:Pzsin_res}
\end{equation}
where Res(n$\Psi_n$) is the resolution of n-th harmonic event plane
defined as $\langle\cos[n(\Psi_n^{\rm obs}-\Psi_n)]\rangle$ (``obs"
denotes an observed angle).

%---------------------------------------------------------------------
\subsection{Feed-down effect}\label{sec:feeddown}

It is known that a significant amount of $\Lambda$ and $\Xi$ hyperons
comes from decays of heavier particles, such as $\Sigma^0$,
$\Sigma^\ast$, and $\Xi$ baryons for $\Lambda$, and $\Xi(1530)$
baryons for $\Xi$.  While the secondary particles from weak decays can
be reduced, though not completely, using information on the decay
topology, particles decayed via strong interaction cannot be separated
from primary particles experimentally due to their short lifetimes.
If the parent particles are polarized, the polarization is transferred
to the daughter hyperons with certain polarization transfer factor, as
discussed with Eqs.~\ref{eq:PlamPxi} and \ref{eq:PlamPom}.  The
transfer factor $C$ depends on the type of decays and could be
negative, for instance, $C_{\Sigma^0\Lambda}=-1/3$ for the
electromagnetic decay of
$\Sigma^0\rightarrow\Lambda+\gamma$~\cite{Becattini:2016gvu}.  Based
on model
studies~\cite{Becattini:2016gvu,Karpenko:2016jyx,Li:2017slc,Xia:2019fjf,Li:2021zwq},
such feed-down contribution is found to suppress the polarization of
inclusively measured \lam compared to primary \lam by $10{\text
  -}20$\% depending on the model used.  In case for $\Xi$ hyperons,
$\Xi(1530)$ has spin $3/2$ and the polarization transfer factor in the
decay of $\Xi(1530)\rightarrow\Xi^{-}+\pi$ is equal to unity.
Therefore the feed-down contribution for $\Xi$ leads to the
enhancement of the polarization of inclusive $\Xi$ by
$\sim$25\%~\cite{Li:2021zwq}.

Although the effect of feed-down is not so significant, it is
important to assess the effect, especially when extracting physical
quantities such as the vorticity and magnetic field at the freeze-out.
Note that the feed-down correction relies on the assumption of local
thermodynamic equilibrium for spin degree of freedom as formulated in
Ref.~\cite{Becattini:2016gvu}.  But it is not clear if the relaxation
time is similar for the vorticity and magnetic field.  Furthermore,
actual situation may be more complicated since some of the particles
have a shorter lifetime than the system lifetime (10--15 fm$/c$).

%----------------------------------------------------------------------
\subsection{Vector mesons spin alignment}
\label{sec:measure-sa}

Vector mesons, $s\!=\!1$ particles, can be also utilized to study the
particle polarization in heavy-ion collisions.  Unlike in the case of
hyperons' weak decay, vector mesons predominantly decay via parity
conserved (strong or electromagnetic) interaction.  Therefore, one
cannot determine the direction of the polarization of vector mesons,
and their polarization state is usually reported via so-called
spin-alignment measurements.  The spin state of a vector meson is
described by the spin density matrix $\rho_{mn}$.  The diagonal
elements of this matrix have a meaning of the probabilities for spin
projections onto a quantization axis to have values $0,~\pm 1$;
$\rho_{00}$ represents the probability for the spin projection to be
zero. As $s_z=\pm 1$ projections can not be distinguished, and the sum
of the probabilities has to be unity, only one independent diagonal
%element, usually $\rho_{00}=w_0$, can be measured. In the case of
element, usually $\rho_{00}$, can be measured. In the case of
vector meson decay into two (pseudo)-scalar mesons, $\rho_{00}$ can be
determined directly from the angular distributions of the vector
mesons decay products (given by the squares of the corresponding
spherical harmonics):
\begin{eqnarray}
%  \frac{dN}{d\cos\theta^\ast} 
%= \frac{3}{4} [1-\rho_{00}+(3\rho_{00}-1)\cos^2\theta^\ast],
  \frac{dN}{d\Omega^\ast} 
= \frac{3}{8\pi} [1-\rho_{00}+(3\rho_{00}-1)\cos^2\theta^\ast],
\label{eq:rho00-1}
\end{eqnarray}
where $\theta^\ast$ is the angle of one of the daughter particles with
respect to the polarization direction in the rest frame of the vector
meson.%~\cite{Liang:2004xn}.
For the global spin alignment measurement, the polarization direction
is given by the orbital angular momentum direction of the system,
perpendicular to the reaction plane.  In the case of unpolarized
particles, $\rho_{00}$ equals $1/3$.  The deviation of $\rho_{00}$
from $1/3$ would indicate spin alignment of vector mesons.

The spin alignment, $\Delta\rho =\rho_{00}-1/3$, can be measured by
directly analyzing $\cos\theta^\ast$ distribution given in
Eq.~\ref{eq:rho00-1}, %or using
or considering $\langle\cos^2\theta^\ast\rangle$ as follows
\begin{eqnarray}
  \langle\cos^2\theta^\ast\rangle
  %&=& \int\!\frac{d\Omega^\ast}{2\pi}
  %\int\!\frac{d\phi_V}{2\pi} \int\!\frac{d\Psi_{\rm RP}}{2\pi} 
  &=& \int\!d\Omega^\ast
\frac{3}{8\pi}\left[1-\rho_{00}+(3\rho_{00}-1)\cos^2\theta^\ast
  \right]
\cos^2\theta^\ast 
%\\ &=&
= \frac{1}{3} +\frac{2}{5}\Delta\rho.
%(1-\rho_{00}) + \frac{1}{5}(3\rho_{00}-1),
\end{eqnarray}
%
%where $\phi_V$ denotes azimuthal angle of vector meson and $\cos\theta^\ast=\sin\theta_D^\ast\sin(\Psi_{\rm RP}-\phi_D^\ast)$
%($D$ indicates daughter product of the vector meson), 
%and \red{it results in}
It results in
\be
\Delta\rho = \frac{5}{2}\left(\mean{\cos^2\theta^\ast}-\frac{1}{3}\right).
\ee
Taking also into account the event plane
resolution~\cite{Tang:2018qtu} one arrives to the equation:
\begin{equation}
\rho_{00} = \frac{1}{3} + \frac{4}{1+3{\rm Res}(2\Psi)} 
\left( \rho_{00}^{obs} - \frac{1}{3} \right),
\label{eq:drhophi}
\end{equation}
where $\rho_{00}^{\rm obs}$ is the measured (``observed'') signal
and Res(2$\Psi$) is the event plane resolution defined as
$\langle\cos[2(\Psi-\Psi_{\rm RP})]\rangle$ where $\Psi$ can be either
the first-order or second-order event plane.  One could also analyze
the daughter product distribution relative to the reaction
plane~\cite{Voloshin:2017kqp,Tang:2018qtu} similarly to that
performed in the global polarization measurement:
\begin{eqnarray}
\rho_{00} &=& \frac{1}{3} - \frac{4}{3}\frac{\langle\cos[2(\phi^\ast-\Psi)]\rangle}{{\rm Res}(2\Psi)},\label{eq:rho00-2}
\end{eqnarray}
where $\phi^\ast$ is the azimuthal angle of the daughter product in the
parent rest frame.

In the case of vector meson decaying into two fermions, e.g. $J/\psi
\rightarrow e^+e^-$, the interpretation of the final angular
distribution in terms of the vector meson polarization is less
straightforward, as it involves the spin wave functions of the
daughter fermions. In this case the angular distribution of the
daughter particles is often parameterized with a set of lambda
parameters. For the distribution integrated over azimuthal angle, it
reduces to
\be
  \frac{dN}{d\cos\theta^\ast} 
\propto \frac{1}{3+\lambda_\theta}[1+\lambda_\theta \cos^2\theta^\ast].
\label{eq:lambdatheta}
\ee
The $\lambda_\theta$ parameter can be then determined from
\be
\mean{\cos^2\theta^\ast}=\frac{1+3\,\lambda_\theta/5}{3+\lambda_\theta}.
\ee
If the masses of the fermions are small, the helicity conservation
tells that they should be in the spin state with projection on their
momentum $\pm 1$. In this case, $\lambda_\theta$
parameter is related to the probability for a vector meson to have
spin projection zero via equation~\cite{Faccioli:2010kd}
\be
%\lambda_\theta =\frac{1-3\, w_0}{1+w_0}.
\lambda_\theta =\frac{1-3\, \rho_{00}}{1+\rho_{00}}.
\ee
%

  % "How is it mwasured"
\subsection{Detector acceptance effects}\label{sec:DetAcc}
%--------------------------------------------------------------
\subsubsection{Polarization along the initial angular momentum}

We start with deriving the correction for polarization measurements
based on Eq.~\ref{eq:PH_nores}.  For the case of an imperfect
detector, one has to take into account that in the calculation of the
average $\langle \sin \left(\psirp-\phi^*\right)\rangle$, the integral
over solid angle $d\Omega^* = d\phi^* \sin \theta^* d \theta^*$ of the
hyperon decay baryon's 3-momentum ${\bf p}^*$ in the hyperon rest
frame, is affected by detector acceptance:
\begin{eqnarray}
\label{meanSinAcc}
\mean{ \sin ( \psirp-\phi^* )} = 
\int
     \frac{d\Omega^*}{4\pi}\frac{d\phi_H}{2\pi} A({\bf p}_H, 
{\bf p}^*) \int\limits_0^{2\pi} \frac{d\psirp}{2\pi} 
\{1+2v_{2,H} \cos[2(\phi_H-\Psi_{\rm RP})]\}
\nonumber
\\ 
\times \sin (\Psi_{\rm RP} -\phi^*)
\left[1+\alpha_H~P_H ({\bf p}_H; \phi_H-\Psi_{\rm RP})\,
\sin \theta^\ast \cdot \sin 
\left(\Psi_{\rm RP} - \phi^\ast \right) \right].~~~~
\end{eqnarray}
Here ${\bf p}_H$ is the hyperon 3-momentum, and $A\left({\bf p}_H,
{\bf p}^*_p\right)$ is a function to account for detector acceptance.
The integral of this function over $(d\Omega^*_p/4\pi)(d\phi_H/2\pi)$
is normalized to unity.  The polarization component along the system 
orbital angular momentum could depend
on the relative azimuthal angle ($\phi_H-\Psi_{\rm RP}$). 
Taking into account the symmetry of the system, one can expand the
polarization as a function of ($\phi_H-\Psi_{\rm RP}$) in a sum over
even harmonics. Keeping below only the first two terms:
\begin{eqnarray}
\label{sumForGlobalPolarization}
P_H\left(\phi_H-\Psi_{\rm RP},p_t^H,\eta^H\right)
= P_0\left(p_t^H,\eta^H\right) +2P_2\left(p_t^H,\eta^H\right)
\cos\left[2(\phi_H-\Psi_{\rm RP})\right].
\end{eqnarray}
Substituting it into Eq.~\ref{meanSinAcc} and integrating over
$\psirp$ one gets
\begin{eqnarray}
\label{GlobalPolarizationSinAcc}
\mean{ \sin ( \psirp-\phi^* )} &=&
\frac{\alpha_H}{2} 
\int \frac{d\Omega^*}{4\pi} 
\frac{d\phi_H}{2\pi}
A\left({\bf p}_H, {\bf p}^*\right) 
\sin\theta^* \nonumber
\\ \nonumber \hspace{2cm} & &
\times \left[(P_0 + 2P_2 v_2) 
- (P_2+ P_0 v_2) 
\cos\left[2(\phi_H-\phi^*)\right] \right]
\\&=&
\frac{\alpha_H\pi}{8}\left[
A_{0}~(P_0+2P_2 v_2) - A_{2}(P_2+P_0 v_2)\right],
\end{eqnarray}
where the ``acceptance'' functions $A_{0}(p_t^H,\eta^H)$ and
$A_{2}(p_t^H,\eta^H)$ are defined by:
\begin{eqnarray}
A_{0}(p_t^H,\eta^H) &=& \ds \frac{4}{\pi} 
\int {\frac{d\Omega^*}{4\pi} 
\frac{d\phi_H}{2\pi}
A\left({\bf p}_H, {\bf p}^*\right) \sin\theta^*}.
\\
\label{AccCoefficientAdditive}
A_{2}(p_t^H,\eta^H) &=& \ds  
\frac{4}{\pi} \int {\frac{d\Omega^*}{4\pi} \frac{d\phi_H}{2\pi}
A\left({\bf p}_H, {\bf p}^*\right) 
\sin\theta^*\cos\left[2(\phi_H-\phi^*)\right]}.
\end{eqnarray}
For the perfect acceptance $A_0=1$ and $A_2=0$.
Similarly one obtains:
\begin{eqnarray}
 & &\mean{ \sin ( \psirp-\phi^* )  \cos[2(\phi_H-\phi^*)] }  
\nonumber
\\ & & \hspace{2cm} 
=
\frac{\alpha_H\pi}{8}\left[
A_{0}~(P_2+P_0 v_2) - \frac{1}{2}A_{2}(P_0 +3 P_2 v_2)\right].
\end{eqnarray}

Another set of equations can be derived for the method based
on calculation of the $\langle\cos\theta^\ast\rangle$, Eq.~\ref{eq:poltheta}.
In this case 
\begin{eqnarray}
\mean{ \sin ( \psirp-\phi^* )\sin\theta^*} &=&
\frac{\alpha_H}{3}\left[
\tilde{A}_{0}~(P_0+2P_2 v_2) - \tilde{A}_{2}(P_2+P_0 v_2)\right],
\end{eqnarray}
\begin{eqnarray}
& & \hspace{0.1cm}\mean{ \sin ( \psirp-\phi^* ) \sin\theta^* \cos[2(\phi_H-\phi^*)] }
\nonumber
\\ & &\hspace{3.5 cm} =
\frac{\alpha_H}{3}\left[
\tilde{A}_{0}~(P_2+P_0 v_2) - \frac{1}{2}\tilde{A}_{2}(P_0 +3 P_2 v_2)\right],
\end{eqnarray}
where
\begin{eqnarray}
\tilde{A}_{0}(p_t^H,\eta^H) &=& \ds \frac{3}{2} 
\int {\frac{d\Omega^*}{4\pi} 
\frac{d\phi_H}{2\pi}
A\left({\bf p}_H, {\bf p}^*\right) \sin^2\theta^*},
\\
\label{AccCoefficientAdditive}
\tilde{A}_{2}(p_t^H,\eta^H) &=& \ds  
\frac{3}{2} \int {\frac{d\Omega^*}{4\pi} \frac{d\phi_H}{2\pi}
A\left({\bf p}_H, {\bf p}^*\right) 
\sin^2\theta^*\cos\left[2(\phi_H-\phi^*)\right]}.
\end{eqnarray}

%--------------------------------------------------------
\subsubsection{Polarization along the beam direction}
\label{sec:acc-Pz}

For $\pz$ measurement, we consider the average of
$\langle\cos\theta_B^\ast\rangle$ using Eq.~\ref{eq:hyperon_decay},
where $\theta_B^\ast$ is the polar angle of the daughter baryon in its
parent hyperon rest frame, relative to the beam direction.
\begin{eqnarray}
 \langle \cos\theta_{\rm B}^\ast \rangle
% &=& \int \frac{dN}{d\Omega^\ast} A\left({\bf p}_H, {\bf p}^\ast\right) \cos\theta_{\rm B}^\ast d\Omega^\ast 
 &=& \int \frac{d\Omega^\ast}{4\pi} A\left({\bf p}_H, {\bf p}^\ast\right)
 (1+\alpha_{\rm H} P_{\rm z}\cos\theta_{\rm B}^\ast) \cos\theta_{\rm
   B}^\ast
 \\
 &=& \alpha_H P_{\rm z} \int \frac{d\Omega^\ast}{4\pi} A\left({\bf p}_H, {\bf p}^\ast\right) \cos^2 \theta_{\rm B}^\ast.
\end{eqnarray}
Thus:
\begin{eqnarray}
  P_z 
%  = \frac{\langle\cos\theta_{\rm B}^\ast\rangle}{\alpha_{\rm H}
%    \int\frac{d\Omega^\ast}{4\pi} \cos^2\theta_{\rm B}^\ast} 
  = \frac{1}{A_{\rm z}}
  \frac{3\langle\cos\theta_{\rm B}^\ast\rangle}{\alpha_{\rm H}},
\end{eqnarray}
where $A_{\rm z}$ is acceptance correction factor defined as
\begin{equation}
  A_{\rm z}(p_t^H,\eta^H)
  = 3\int \frac{d\Omega^\ast}{4\pi}
  A\left({\bf p}_H, {\bf p}^\ast\right) \cos^2\theta_{\rm B}^\ast.
  \label{eq:Az}
\end{equation}
The factor $A_z$ can be determined in a data driven way, similar to
the acceptance correction factors in the global polarization
measurement, and is typically close to unity~\cite{Adam:2019srw}.

%-----------------------------------------------------------
\subsubsection{Acceptance effects in spin alignment measurements}
\label{sec:accspinalign}

Spin alignment measurements are significantly more difficult compared
to the measurements of the hyperon polarization. The difficulty comes
from the fact that while the acceptance effects in the polarization
measurements can only change the magnitude of the effect, in the spin
alignment measurement the acceptance effects could lead to false
spurious signal.  We demonstrate this below providing equations for
the acceptance correction to the signal for the case of vector mesons
experiencing elliptic flow.

One of the main tracking efficiency effects is due to different
probabilities if vector meson reconstruction when the daughter
particles are emitted along the momentum of the parent particles or
perpendicular to that. 
%This simulation studies 
A toy model study on decay daughters can show that such efficiency
effect can be well parameterized by parameter $a_2$ in the equation
\be
A(\phist)=A_0(1+2a_2(\pt)\cos[2(\phist - \phi)],
\ee
where $\phi$ denotes the vector meson azimuthal angle in the
laboratory frame.

Then following Eq.~\ref{eq:drhophi} and accounting for elliptic flow
and efficiency effects, one finds:
\bea \Delta \rho^{\rm obs} &\equiv& \rho_{00}^{\rm obs}
-\frac{1}{3}=\int \frac{d\Psi_{\rm RP}}{2\pi} \frac{d\phist}{2\pi}
\frac{d\phi}{2\pi} (-4/3)\cos[2(\phist -\psirp)]
\nonumber \\ &\times&
(1+2 v_2(\pt) \cos[2(\phi-\psirp)]) (1+2 a_2(\pt)
\cos[2(\phist-\phi)])
\nonumber \\ &\times&
(1-\frac{3}{2}\Delta\rho(\pt) \cos[2(\phist-\psirp)])
=\Delta\rho-\frac{4}{3} a_2 v_2,
\label{eq:drhoobs}
\eea
where the superscript ``obs''
(observed) denotes the value obtained by the direct application of
Eq.~\ref{eq:drhophi} to the data. One can see that even in the case of
``real'' $\Delta\rho$ being zero, the observed signal is not zero due
to interplay of the elliptic flow and tracking efficiency effects.

Similarly this effect biases the flow measurement as:
\be
v_2^{\rm obs} = v_2-\frac{3}{4}a_2 \Delta\rho,
\label{eq:v2obs}
\ee
and the determination of parameter $a_2$ from data can be done with:
\be
a_2^{\rm obs} = a_2-\frac{4}{3} v_2 \Delta\rho.
\label{eq:a2obs}
\ee
The actual values of the spin alignment and elliptic flow can be
obtained by solving the above equations
Eqs.~\ref{eq:drhoobs}--\ref{eq:a2obs} with respect to $v_2$, $a_2$,
and $\Delta\rho$.

The equations above demonstrate only one example of the tracking
efficiency effects leading to a spurious spin alignment signal.
Another example was discussed in Ref.~\cite{Lan:2017nye}, where the
authors investigated (and found to be significant) the effect of the
finite rapidity acceptance on $\Delta\rho$ measurements.

%\end{widetext}

%

%========================
\section{Overview of experimental results\label{sec:result}}
%==================================================================
\subsection{Global polarization of $\Lambda$ hyperons}\label{sec:result-global}

\subsubsection{Energy dependence}

\begin{figure}[th]
\centerline{\includegraphics[width=0.7\linewidth]{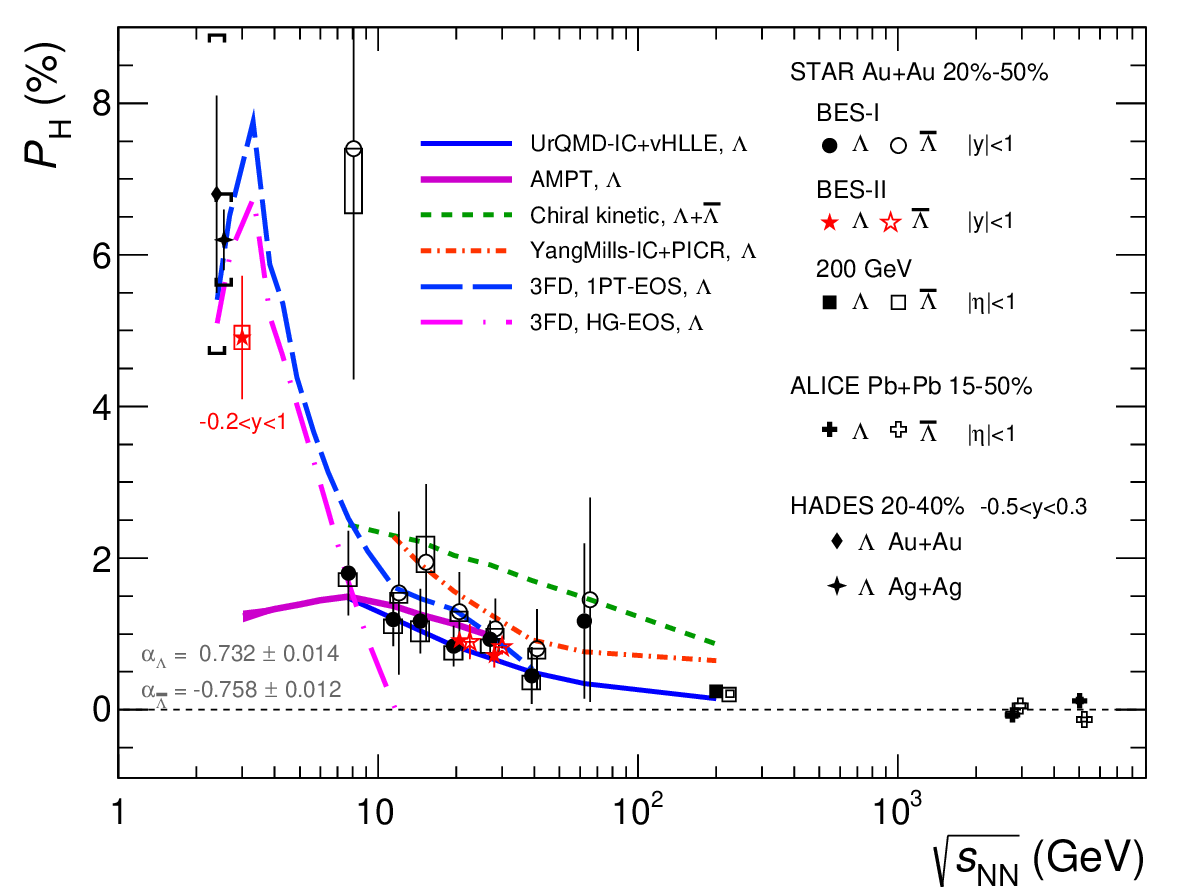}}
\caption{Collision energy dependence of \lam and \alam global
  polarization for mid-central heavy-ion
  collisions~\cite{STAR:2021beb} compared to various model
  calculations~\cite{Karpenko:2016jyx,Li:2017slc,Sun:2017xhx,Xie:2017upb,Ivanov:2020udj}. The
  experimental data from the original publications are rescaled
  accounting for the recent update of the \lam decay
  parameters~\cite{Zyla:2020zbs} indicated in the figure.}
\label{fig:PHvsRootS-lam}
\end{figure}

Global polarization of \lam and \alam hyperons has been measured in a
wide range of collision energies. The first observation of non-zero
global polarization was reported in Au+Au collisions at \snn = 7.7--39
GeV in the first phase of the beam energy scan program (BES-I) at RHIC
by the STAR Collaboration~\cite{STAR:2017ckg}; later it was also
confirmed with a better significance at \snn =
200~GeV~\cite{Adam:2018ivw}.  At the LHC energies the measurements
were performed by the ALICE Collaboration~\cite{Acharya:2019ryw}.
Figure~\ref{fig:PHvsRootS-lam} presents a compilation of the results
of the global polarization measurements for \lam and \alam hyperons at
mid-rapidity for mid-central collisions as a function of collision
energy.  The polarization increases as the collision energy
decreases. One would naively expect that the initial orbital angular
momentum becomes larger at higher energy~\cite{Karpenko:2016jyx},
therefore the polarization would have the same trend, but this
argument does not take into account that the initial angular momentum
has to be spread over much larger rapidity region and more produced
particles, and that the particle production at the midrapidity is
almost boost invariant.  Another reason for the observed energy
dependence might be a dilution effect of the vorticity due to longer
lifetime of the system at higher collision energies.

Most of the theoretical calculations rely on the assumptions that (a)
the system is in a local thermal equilibrium and (b) that the spin
polarization is not modified at later non-equilibrium stages, e.g., by
hadronic rescattering~\cite{Barros:2005cy,Barros:2007pt}. Neither of
these assumptions is obvious.  Nevertheless, most of the calculations,
based on different approaches, such as hydrodynamic
models~\cite{Karpenko:2016jyx,Xie:2017upb,Ivanov:2019ern,Ivanov:2020udj},
chiral kinetic approach~\cite{Sun:2017xhx}, and a transport
model~\cite{Li:2017slc}, surprisingly well reproduce the observed
energy dependence of the global polarization at the quantitative
level, as seen in Fig.~\ref{fig:PHvsRootS-lam}. Note that there still
exists a disagreement between the data and models in differential
measurements, which we discuss in the following sections.  Based on
Eq.~\eqref{eq:polar}, the vorticity can be estimated as $\omega\approx
k_BT(P_{\lam}+P_{\alam})/\hbar$ with $T$ being the system temperature
at the time of particle emission.  The polarization averaged over \snn
in the BES-I results in $\omega\approx(9\pm1)\times10^{21}~{\rm
  s}^{-1}$, leading to the finding of the most vortical fluid ever
observed~\cite{STAR:2017ckg}.

From empirical estimates~\cite{Voloshin:2017kqp} based on the directed
flow measurements, see Sec.~\ref{sec:estimates}, the global
polarization signal at the LHC energies is expected to be an order of
a few per mill. The results from the ALICE Collaboration are
consistent with zero with statistical uncertainties of the order of
the expected signal. At lower energies, it is expected that the
kinematic vorticity becomes maximum around \snn = 3~GeV and vanishes
at \snn = $2m_{N}$ ($m_{N}$ is the nucleon mass) near the threshold of
nucleon pair production because the total angular momentum of the
system at such energies becomes close to
zero~\cite{Deng:2020ygd,Ayala:2021xrn,Guo:2021udq}.  In such high
baryon density region, the system would no longer experience a
partonic phase but be in a hadronic phase during the entire system
evolution.  Therefore, it would be interesting to check whether the
polarization changes smoothly with the beam energy.  Recently the STAR
Collaboration has reported \lam global polarization in Au+Au
collisions at \snn = 3~GeV~\cite{STAR:2021beb}, followed by results on
\lam global polarization in Au+Au collisions at \snn = 2.4~GeV and
Ag+Ag collisions at \snn = 2.55~GeV by the HADES
Collaboration~\cite{HADES:2022enx}.  The results indicate that the
global polarization still increase at these energies, although the
current uncertainties may be too large to see the expected trend.

Calculation from the three-fluid dynamics (3FD)~\cite{Ivanov:2020udj}
incorporating the equation of state (EoS) for the first-order phase
transition (1PT) captures the trend of the experimental data. The 3FD
model also shows sensitivity of the global polarization to EoS as seen
in some difference in the calculations for the first-order phase
transition and hadronic (HG) EoS.

%-----------------------------------------------------------------------
\subsubsection{Particle-antiparticle difference}

\begin{figure}[th]
  \centerline{\includegraphics[width=0.7\linewidth,bb=0 0 250 213]{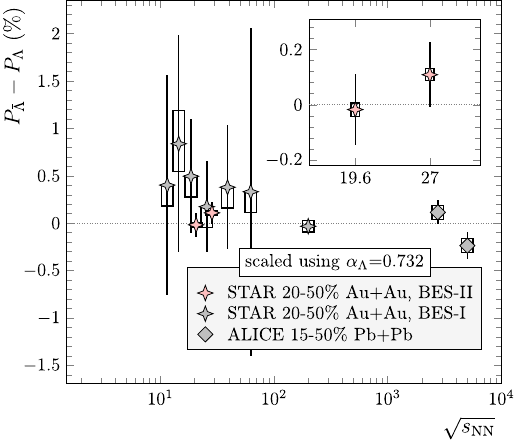}}
\caption{Collision energy dependence of the difference in global
  polarizations of \alam and \lam hyperons,
  $P_{\bar{\Lambda}}-P_{\Lambda}$.  The figure is taken from
  Ref.~\cite{STAR:2023nvo}. }
\label{fig:dPH}
\end{figure}

As discussed in Sec.~\ref{sec:Bfield}, the initial and/or later-stage
magnetic field created in heavy-ion collisions could lead to a
difference in the global polarizations of particles and antiparticles.
The experimental results, presented in Fig~\ref{fig:PHvsRootS-lam}, do
not show any significant difference in polarizations of \lam and
\alam, already indicating that the thermal vorticity, rather than the
magnetic field contribution, is the dominant source of the observed
global polarization.  Figure~\ref{fig:dPH} presents directly the
differences in the global polarizations of \lam and \alam as a
function of \snn~\cite{STAR:2023nvo}. The new RHIC BES-II results from
Au+Au collisions at 19.6~GeV and 27~GeV greatly improve the statistical
uncertainty in the measurements, and show no significant difference
between particle-antiparticle polarizations.  Following
Eq.~\ref{eq:polar}, one could put an upper limit on the magnetic field
effect assuming the local thermodynamic equilibrium for the spin
degrees of freedom:
\begin{equation}
\Delta P_H = P_{\bar{\Lambda}} - P_{\Lambda} = \frac{2|\mu_{\Lambda}|B}{T},
\end{equation}
where $\mu_{\Lambda}=-\mu_{\bar{\Lambda}}=-0.613\mu_N$ with $\mu_N$
being the nuclear magneton.  Thus, one arrives at the upper limit on
the magnitude of the magnetic field $B\lesssim 10^{13}$~T assuming the
temperature $T=150$~MeV and ignoring the feed-down contributions (see
Sec.~\ref{sec:feeddown}).  The estimated magnitude of the magnetic
field is still considerably large and presents an important input for
the dynamical modeling e.g., magneto-hydrodynamics, constraining the
electric conductivity of the plasma.

It should be noted that several other sources could contribute to the
polarization difference.  Ref.~\cite{Vitiuk:2019rfv} suggests that the
different space-time distributions and emission times of \lam and
\alam hyperons lead to the polarization difference.  \alam hyperons,
emitted earlier in time, are less affected by the dilution of the
vorticity with the system expansion, leading to larger polarization of
\alam.  On the other hand, Ref.~\cite{Guo:2019joy} argues that the
formation time of \lam is smaller than that of \alam, leading to
larger polarization of \lam.  The actual situation might be even more
complicated since the spin-orbit coupling may take place at quark
level.  Ref.~\cite{Csernai:2018yok} reported that the strong
interaction with meson field could make \alam polarization larger.
The effect of chemical potential becomes important at lower energies
as it appears in the Fermi-Dirac distribution (see Eq.~\eqref{basic}).  
Non-zero baryon chemical potential is expected to lead to larger polarization of
\alam~\cite{Fang:2016vpj}, though the effect may be rather small. The
feed-down effects with non-zero baryon chemical potential might lead
to the opposite relation~\cite{Becattini:2016gvu} but it would depend
on the relative abundance at different phase space.
Having these complications in mind, non-significant difference in the observed
global polarization of \lam and \alam, $P_{\alam}-P_{\lam}$, does not exclude
a limited contribution from the magnetic field.

%------------------------------------------------------
\subsubsection{Differential measurements}

\begin{figure}[th]
\includegraphics[width=0.48\linewidth]{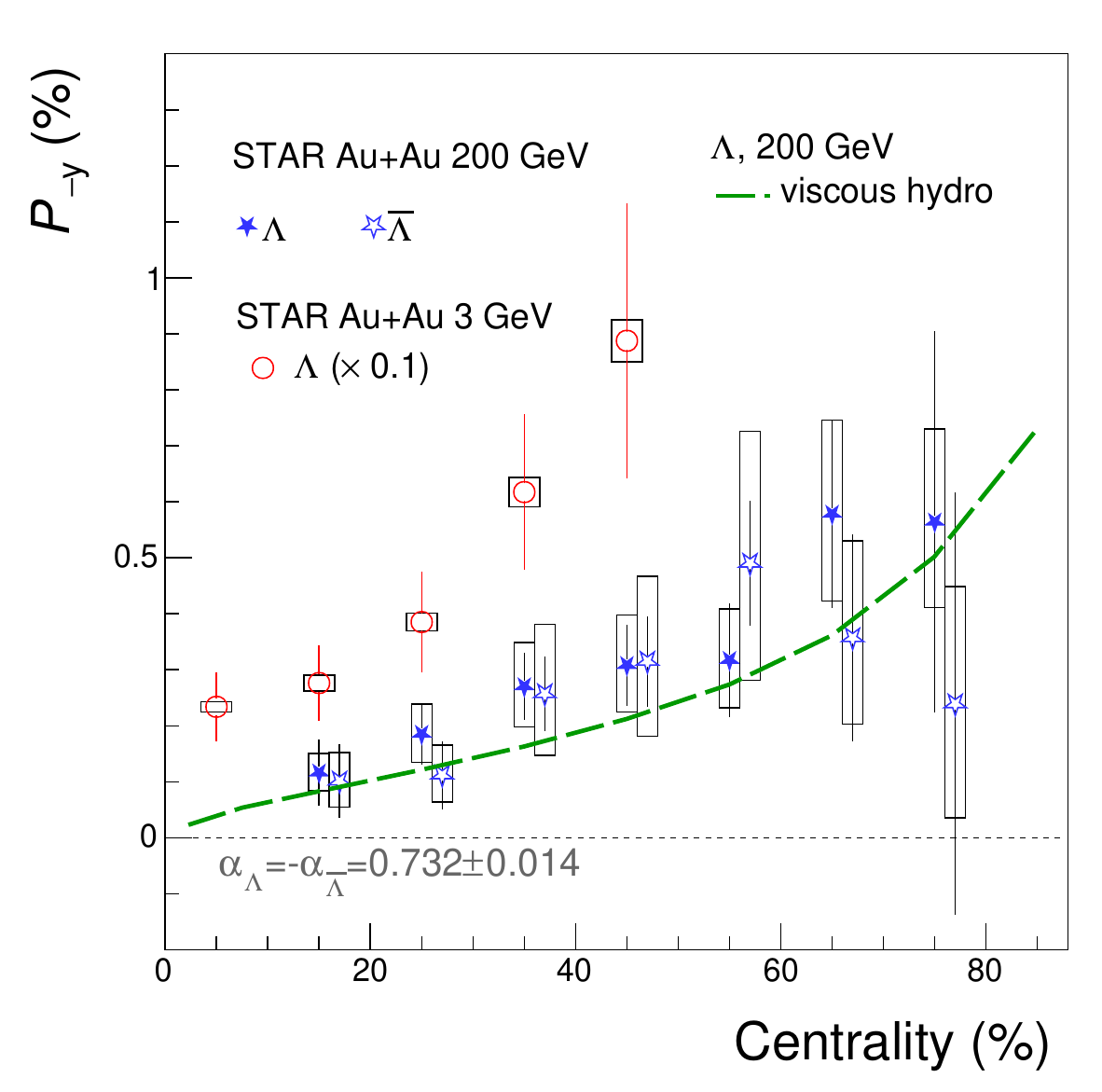}
\includegraphics[width=0.48\linewidth]{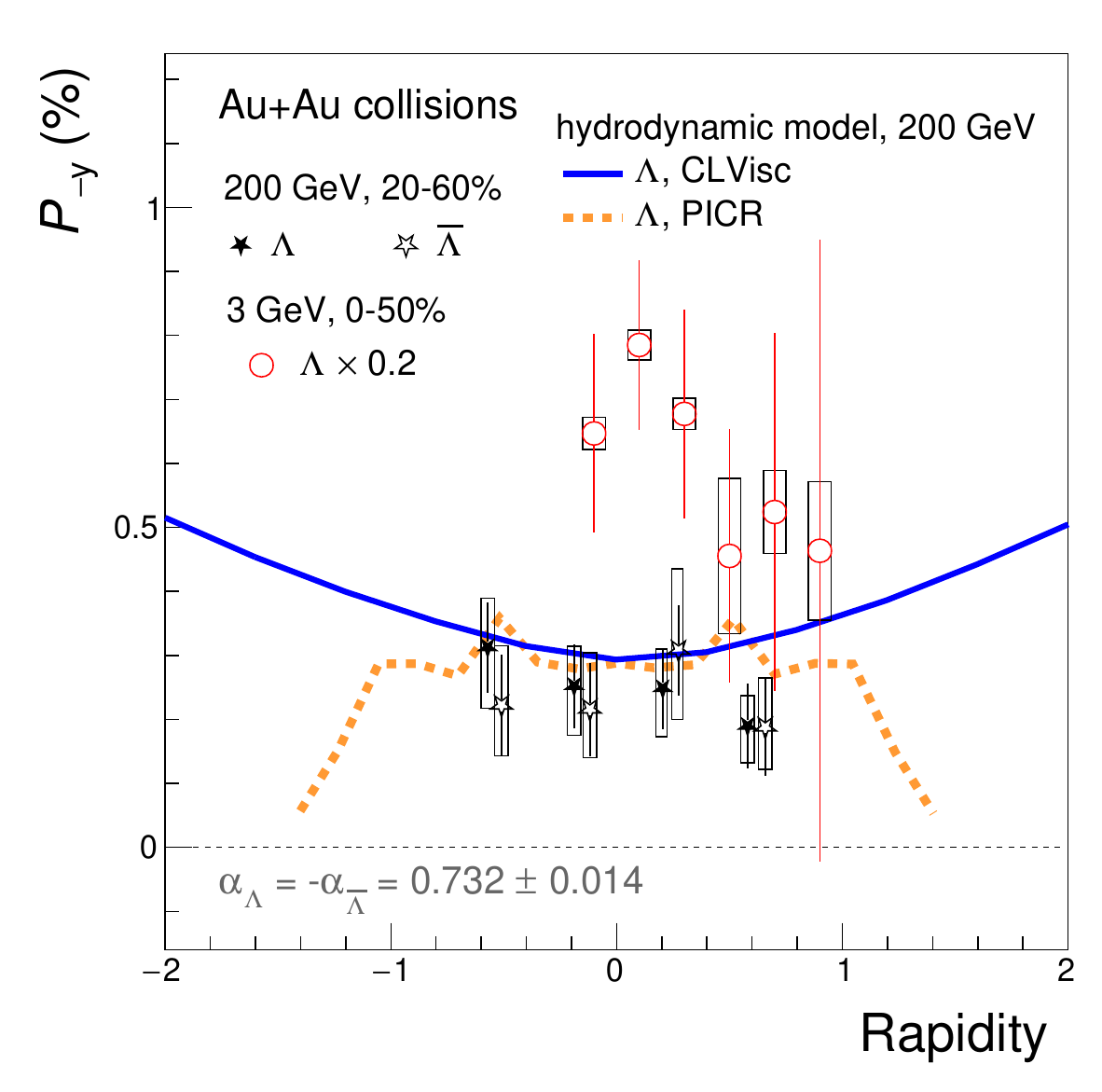}
\caption{(Left) Centrality dependence of \lam\!(\alam) of $\pmy$
  polarization component in Au+Au collisions at \snn = 3, and 200~GeV
  compared to viscous hydrodynamic model
  calculation~\cite{Ryu:2021lnx}.  (Right) Rapidity dependence of
  \lam\!(\alam) $\pmy$ compared to Particle-in-Cell Relativistic
  (PICR) hydrodynamics model~\cite{Xie:2019jun} and viscous
  hydrodynamic model CLVisc~\cite{Wu:2019eyi}.  Note that the data for
  3~GeV in the left (right) plot are scaled by 0.1 (0.2), and the
  average pseudorapidity for 200~GeV is converted to the rapidity in
  the right panel.  }
\label{fig:PHvsCent}
\end{figure}

% Centrality depependence
Recently available high statistics data permit to study global 
polarization differentially, as a function of centrality, transverse
momentum, and rapidity.  Model calculations show that the initial
angular momentum of the system increases from central to mid-central
collisions and then decreases in peripheral collisions since the
energy density decreases~\cite{Becattini:2007sr}, but the vorticity,
hence the global polarization, are expected to increase in more
peripheral collisions~\cite{Jiang:2016woz}.
Figure~\ref{fig:PHvsCent}(left) shows centrality dependence of \lam
(\alam) global polarization in Au+Au collisions at \snn~=~200 and
3~GeV~\cite{Adam:2018ivw,STAR:2021beb}, where the increasing trend
towards peripheral collisions can be clearly seen.  Viscous
hydrodynamics models~\cite{Xie:2017upb,Ryu:2021lnx} qualitatively
describe the centrality dependence of global polarization as shown in
the figure.

% Rapidity dependence
As already mentioned, the ``global'' polarization refers to the
polarization component along the system orbital angular momentum
averaged over all particles and all momenta.  The same component
(denoted as $\pmy$), but measured for a particular kinematics, can
deviate from the global average; in this case the term ``local''
polarization is more appropriate.  For example, the initial velocity
shear resulting in the global vorticity would change with rapidity,
i.e., the shear might be larger in forward/backward rapidity, also
depending on the collision energy ~\cite{Deng:2016gyh,Jiang:2016woz}.
Theoretical models such as hydrodynamics and transport models predict
the rapidity dependence
differently~\cite{Wei:2018zfb,Wu:2019eyi,Xie:2019jun,Liang:2019pst,Guo:2021udq};
some models predict that the polarization goes up in forward
(backward) rapidity while the others predict decreasing trend in
larger rapidities.  The hydrodynamic models using different initial
conditions and frameworks also predict different trends (see
Fig.~\ref{fig:PHvsCent}(right)).  The first study was performed at
\snn = 200~GeV~\cite{Adam:2018ivw} as shown in
Fig.~\ref{fig:PHvsCent}(right) and no significant rapidity dependence
was observed, which may be expected at high collision energy as the
shear should be weaker at midrapidity because of longitudinally boost
invariance.  Recent measurement at \snn = 3~GeV from
STAR~\cite{STAR:2021beb} also found no strong rapidity dependence
within $-0.2<y<1$, even at the rapidity close to the beam rapidity
($y_{\rm beam}=1.02$ at \snn = 3~GeV). Similarly, no dependence on
rapidity is observed at \snn = 2.55~GeV by the HADES
experiment~\cite{HADES:2022enx}.  The uncertainties of the data are
still large and this question should be further studied in future
analyses with better statistics and upgraded/new detectors.

% Figs: centrality dependence and rapidity dependence

\begin{figure}[t]
\begin{center}
\begin{minipage}[b]{0.48\linewidth}
\includegraphics[width=\linewidth]{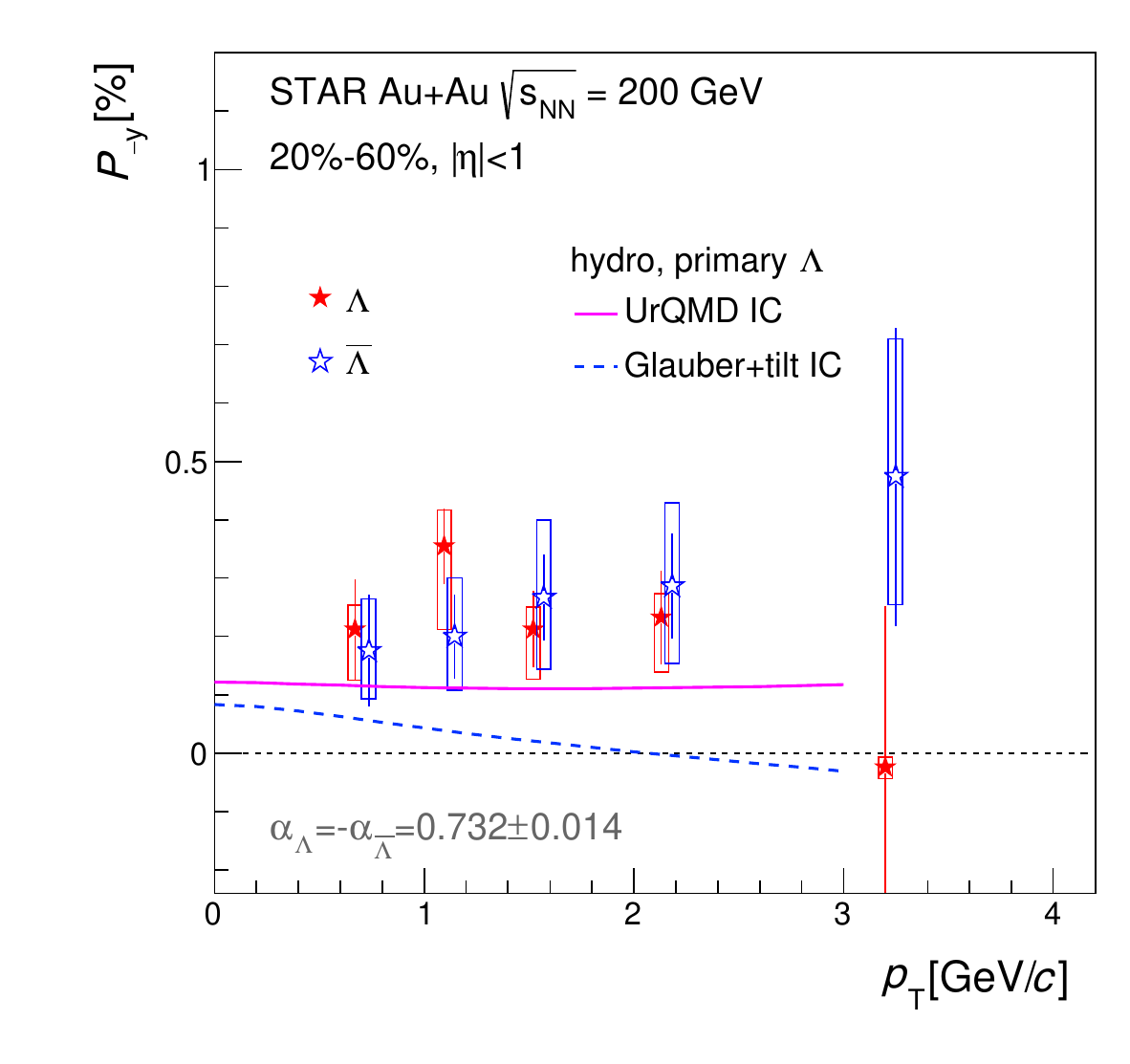}
\end{minipage}
\begin{minipage}[b]{0.50\linewidth}
\includegraphics[width=\linewidth]{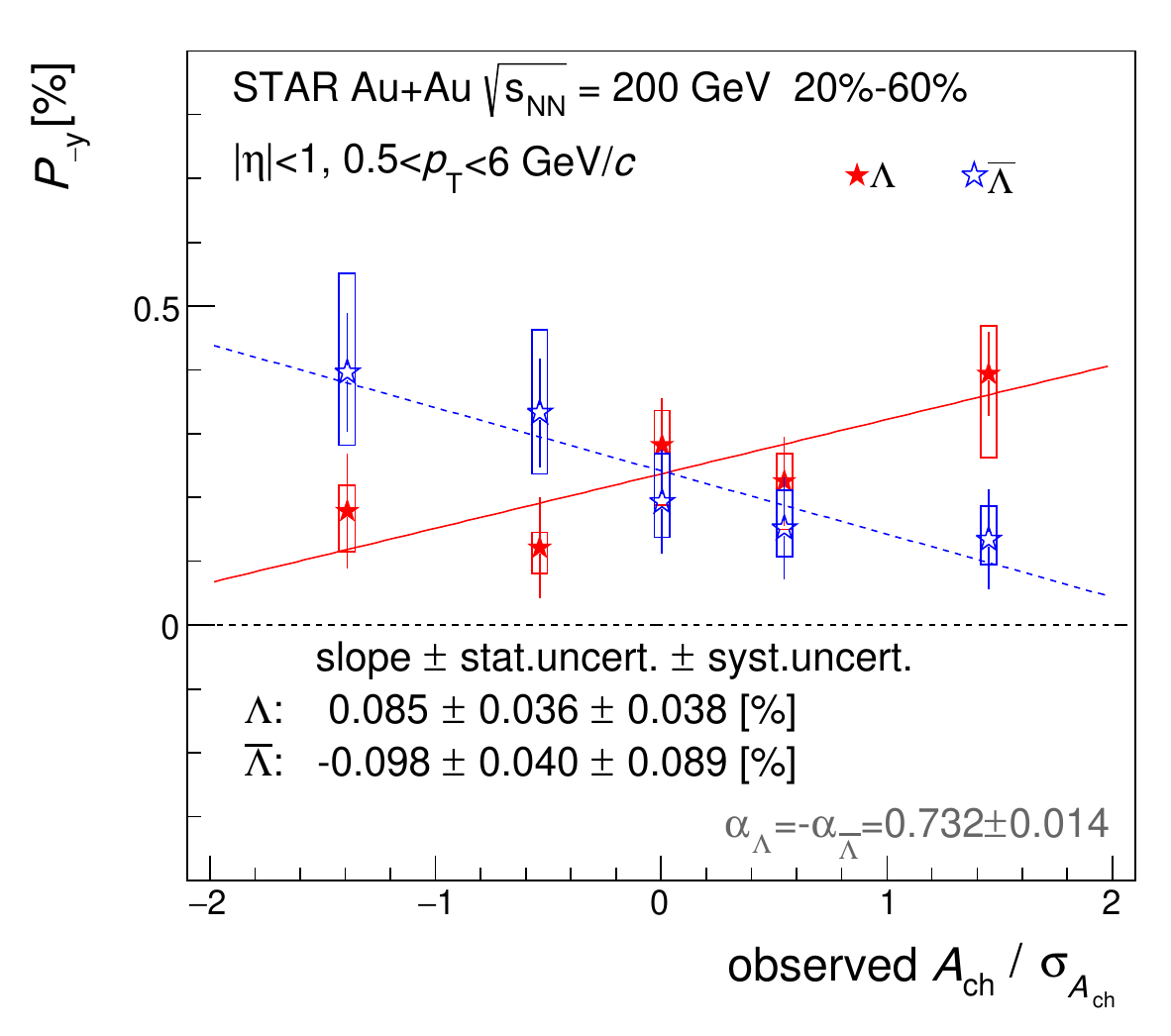}
\vspace{-0.35 cm}
\end{minipage}
\caption{$\pmy$ polarization component of \lam and \alam 
	as a function of (left) transverse momentum dependence and (right)
  charge asymmetry $A_{\rm ch}$ normalized with its RMS in Au+Au collisions
  at \snn = 200~GeV. The figures are adapted from Ref.~\cite{Adam:2018ivw}.  }
\label{fig:PHvsPtAch}
\end{center}
\end{figure}

% pT dependence
It should be noted that the polarization $\pmy$ component seems to
have little dependence on the hyperon transverse momentum
$p_T$~\cite{Adam:2018ivw,STAR:2021beb,HADES:2022enx,STAR:2023nvo},
which qualitatively agrees with theoretical models that predict a mild
$p_T$ dependence. Figure~\ref{fig:PHvsPtAch}(left) shows hyperons'
transverse momentum dependence of the polarization along the system
angular momentum in Au+Au collisions at \snn = 200~GeV, compared to
hydrodynamic model calculations with two different initial
conditions~\cite{Becattini:2016gvu}: Monte Carlo Glauber with the
initial source tilt and UrQMD initial state.  The UrQMD initial
condition includes the initial flow from a preequilibrium phase that
would affect the initial velocity field. Similar trend was also seen
at lower collision
energies~\cite{STAR:2021beb,HADES:2022enx,STAR:2023nvo}.

% charge asymmetry
The STAR Collaboration also studied charge asymmetry ($A_{\rm ch}$)
dependence of the global polarization for a possible relation to
anomalous chiral effects~\cite{Kharzeev:2015znc}. According to
Ref.~\cite{Baznat:2017jfj}, the global polarization could be explained
by axial charge separation due to the chiral vortical effect. In
addition, the axial current ${\bm J}_5$ can be generated in the system
with nonzero vector chemical potential $\mu_{\rm v}$ under a strong
magnetic field ${\bm B}$ (${\bm J}_5\propto Qe\mu_{\rm v}{\bm B}$),
{\it aka} chiral separation effect, where $Qe$ represents net electric
charge of particles. For massless quarks, their momentum direction is
aligned (anti-aligned) with spin direction for right-handed
(left-handed) quarks.  Thus the ${\bm J}_5$, if generated, might
contribute to the hyperon global polarization.  The event charge
asymmetry defined as $A_{\rm ch}=(N_+-N_-)/(N_++N_-)$ where $N_+(N_-)$
is the number of positively (negatively) charged particles was used to
study the possible relation with the polarization assuming $A_{\rm
  ch}\propto \mu_{\rm v}$.  Figure~\ref{fig:PHvsPtAch}(right) shows
\lam and \alam global polarization as a function of $A_{\rm ch}$ for
mid-central Au+Au collisions at \snn = 200~GeV. There seems a slight
dependence on $A_{\rm ch}$ and the slopes look different for \lam and
\alam, although the effect is only at $\sim2\sigma$ level.  The effect
of the chemical potential may be an alternative explanation of the
difference if the charge asymmetry is correlated with the baryon
number asymmetry~\cite{Fang:2016vpj,Ko:2020int}.

\begin{figure}[t]
\begin{center}
\includegraphics[width=0.60\linewidth]{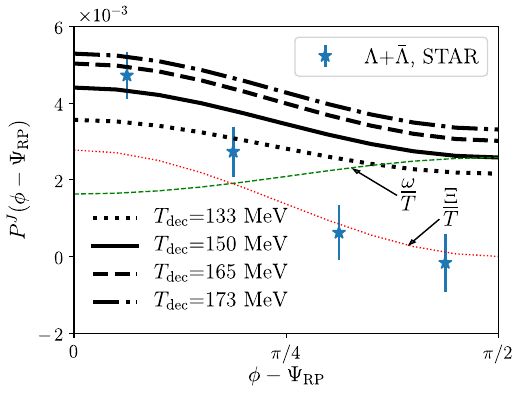}\vspace{-0.1cm}
\caption{Polarization of \lam and \alam hyperons along the initial
  angular momentum $P_J=\pmy$ as a function of hyperons' azimuthal angle relative
  to the second-order event plane $\Psi_{2}$ in 20-50\% Au+Au
  collisions at \snn = 200~GeV (preliminary result from
  STAR~\cite{Niida:2018hfw}), comparing to the hydrodynamic model
  (vHLLE for 20-60\% Au+Au collisions)~\cite{Becattini:2021iol} where
  $T_{\rm dec}$ is a decoupling temperature assuming the isothermal
  freeze-out.  This figure is taken from
  Ref.~\cite{Becattini:2021iol}.  }
\label{fig:PHvsPsi2}
\end{center}
\end{figure}

% phi dependence
Azimuthal angle dependence of the polarization is also of great
interest and has been the subject of debate.  The experimental
preliminary result from STAR~\cite{Niida:2018hfw} shows larger
polarization for hyperons emitted in the in-plane direction than those
in out-of-plane direction as shown in Fig.~\ref{fig:PHvsPsi2}, while
hydrodynamic and transport models predict it oppositely, i.e., larger
polarization in out-of-plane
direction~\cite{Becattini:2015ska,Xie:2017upb,Wei:2018zfb,Fu:2020oxj}.
Based on Glauber simulation shown in Fig.~\ref{fig:glauber}(b), one
expects $\omega_J (\propto dv_z/dx)$ to be larger in in-plane
direction ($x$-direction in the plot), which is consistent with
experimental results.  As shown in Fig.~\ref{fig:PHvsPsi2}, the
calculation including only the contribution from the kinematic
vorticity leads to the opposite sign, while the inclusion of the shear
term leads to the correct sign.  We discuss this question further in
Sec.~\ref{sec:result-Pz} together with the results on polarization
along the beam direction in relation to the so-called ``spin sign
crisis''.

%--------------------------------------------------------------
\subsection{Global polarization of multi-strange hyperons}

\begin{figure}[th]
\centerline{\includegraphics[width=0.7\linewidth]{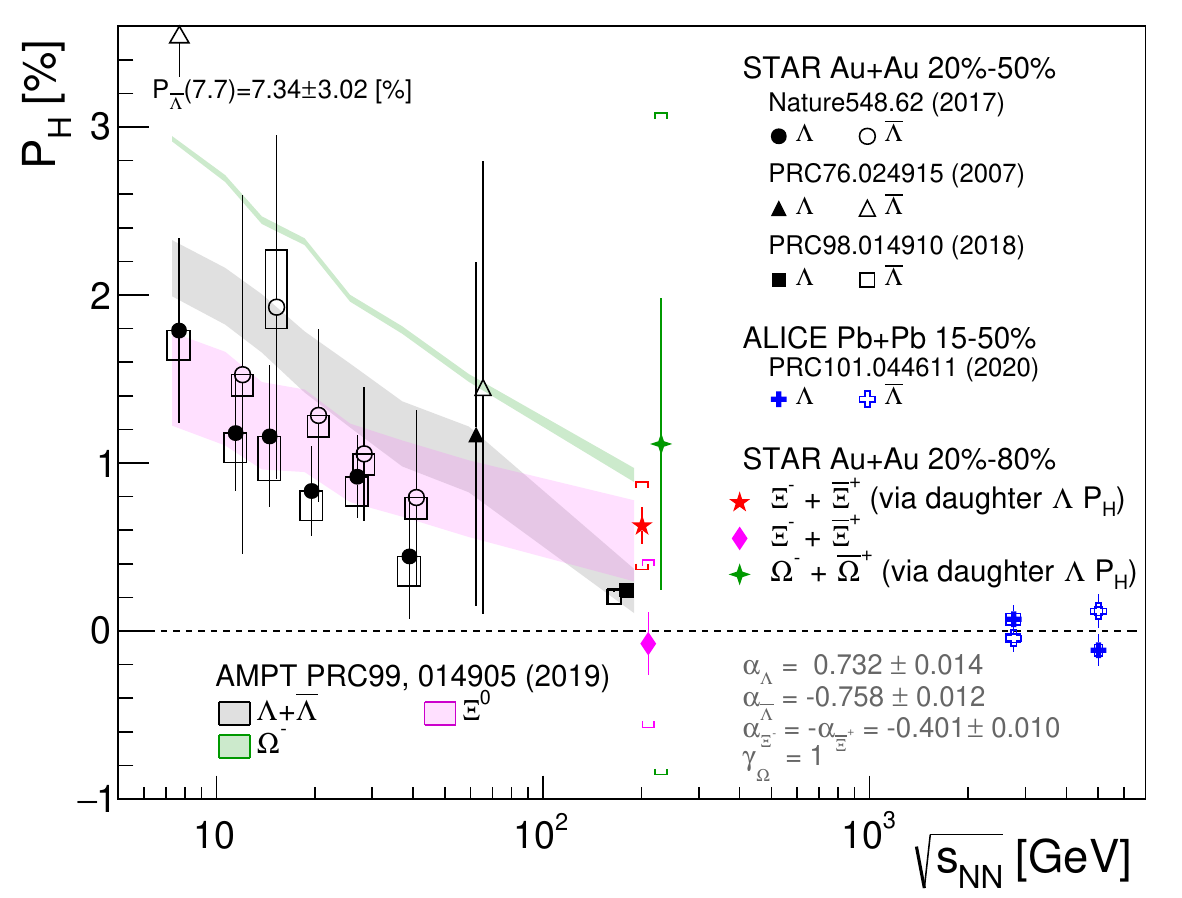}}
\caption{Global polarization of $\Xi$ and $\Omega$ hyperons compared
  to that of \lam\!(\alam) as well as transport model
  calculations. The figure is taken from Ref.~\cite{Adam:2020pti}.}
\label{fig:PHxi}
\end{figure}

Based on the picture of the rotating system, any non-zero spin
particles should be polarized in a similar way, along the direction of
the initial orbital angular momentum.  According to
Eq.~\eqref{eq:polar}, the magnitude of the polarization depends on the
spin of particles. Thus, it is of great interest to study the
polarization of different particles with different spin.
The STAR Collaboration reported global polarization of \xin(\xibar)
and \om(\ombar) hyperons in 200 GeV Au+Au collisions, see
Fig.~\ref{fig:PHxi}. Two independent methods (see
Sec.~\ref{sec:multistrange}) were used to measure \xin(\xibar)
polarization and the results combining \xin and \xibar, and averaging
over the two methods is found to be positive at the 2$\sigma$ level
($\langle P_{\Xi}\rangle=0.47\pm0.10({\rm stat})\pm0.23({\rm syst})\%$
for 20-80\% centrality), supporting the global vorticity picture.  The
cascade polarization is measured to be slightly larger than that of
inclusive \lam, but the significance of that is below 1$\sigma$.  The
results on \om global polarization hint even larger polarization
indicating a possible hierarchy of $P_{\Omega} > P_{\Xi} >
P_{\Lambda}$ but with large uncertainties.  Based on
Eq.~\eqref{eq:polar}, the following relation: $P_{\Lambda} = P_{\Xi^-}
= \frac{3}{5}P_{\Omega^-}$ is expected.  Recent model study shows that
this relation is valid only for primary particles, while it leads to
$P_{\Lambda} < P_{\Xi^-} < P_{\Omega^-}$ after taking into account the
feed-down contribution~\cite{Li:2021zwq}, which seems to be consistent
with the data.  More precise measurements are needed to clarify
the particle/spin dependence of the global polarization.

It is worth mentioning that \om hyperon has larger magnetic moment
($\mu_{\Omega^-}=-2.02\mu_{\rm N}$) compared to those for \lam
($\mu_{\Lambda}=-0.613\mu_{\rm N}$) and \xin
($\mu_{\Xi^-}=-0.65\mu_{\rm N}$).  Therefore, the polarization
difference between \om and \ombar, if any, should be more sensitive to
the magnetic field created in the collisions.  Another thing to be
mentioned is that one of the decay parameter $\gamma_{\Omega}$ is
unknown, but expected to be close to either $+1$ or $-1$ (see
Sec.~\ref{sec:multistrange}).  Assuming the vorticity picture, one can
determine the sign of $\gamma_{\Omega}$.  Currently the experimental
result on $\Omega$ global polarization has large uncertainty but
future high statistics data will allow to resolve the ambiguity.

%---------------------------------------------------------------
\subsection{Global spin alignment of vector mesons}

\begin{figure}[th]
\centerline{\includegraphics[width=0.80\linewidth]{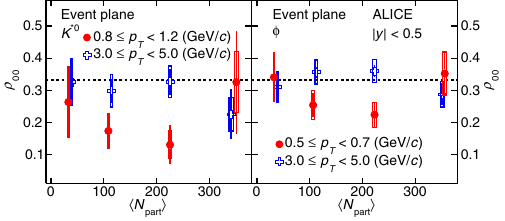}}
\caption{Global spin alignment of $K^{\ast0}$ and $\phi$ mesons shown
  as the spin density matrix element $\rho_{00}$ with quantization
  axis chosen along the system orbital angular momentum measured in
  Pb+Pb collisions at \snn=2.76~TeV by the ALICE Collaboration.  The
  figure is taken from Ref.~\cite{ALICE:2019aid}.}
\label{fig:rho00-lhc}
\end{figure}

\begin{figure}[th]
\begin{tabular}{cc}
  \begin{minipage}[t]{0.48\linewidth}
    \centering
    \includegraphics[width=\linewidth]{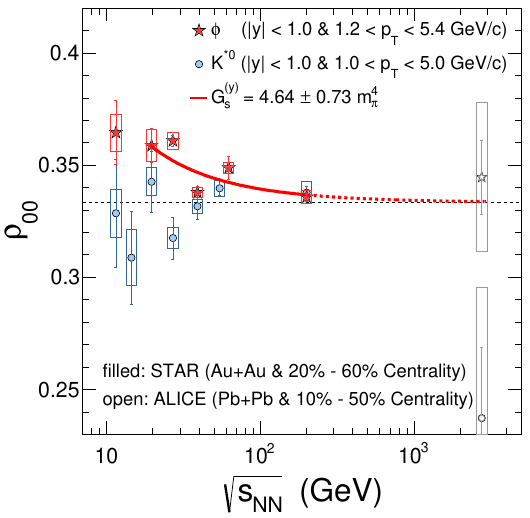}
  \end{minipage} &
  \begin{minipage}[t]{0.48\linewidth}
    \centering
    \includegraphics[width=\linewidth]{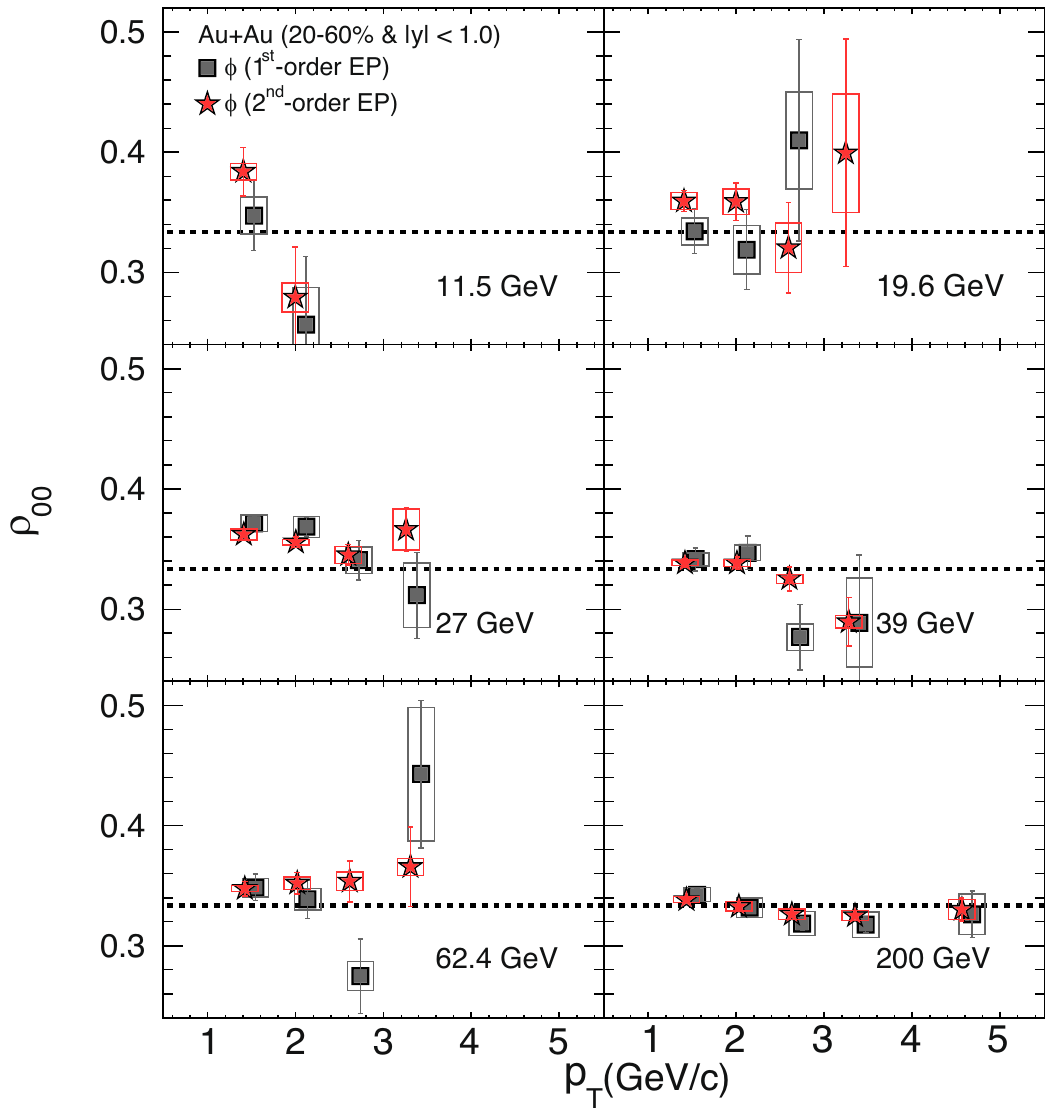}
  \end{minipage}\\
  \end{tabular}

  \vspace{0.5cm}
  \begin{minipage}[t]{\linewidth}
    \centering
    \includegraphics[width=0.98\linewidth]{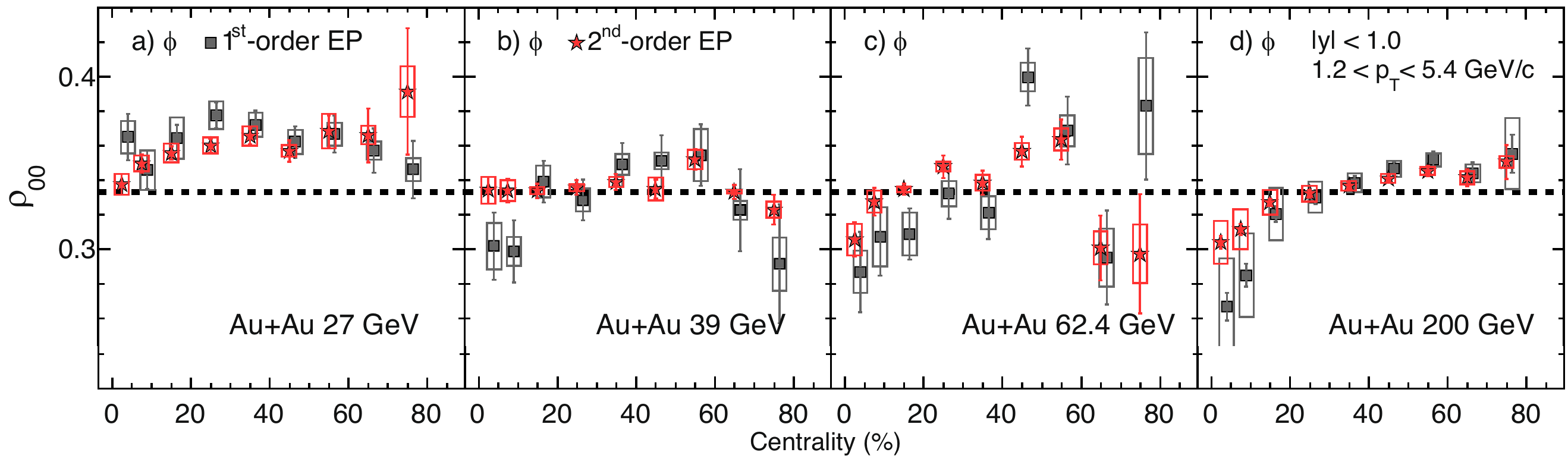}
  \end{minipage}
  \vspace{-0.5cm}
  \caption{$K^{\ast0}$ and $\phi$ mesons spin density matrix element
    with quantization axis along the system orbital angular momentum
    in Au+Au collisions by the STAR experiment. The results are shown
    as a function of collision energy \snn, transverse momentum, and
    centrality.  The figures are taken from
    Ref.~\cite{STAR:2022fan}.}
\label{fig:rho00-rhic}
\end{figure}

The vorticity should also lead to the global polarization of the
vector mesons, such as $K^{\ast0}$ and $\phi$, revealing itself via
global spin alignment~\cite{Voloshin:2004ha,Liang:2004xn}.  The first
measurement of the spin alignment was made by the STAR Collaboration
at RHIC using 200~GeV Au+Au collisions in 2008~\cite{Abelev:2008ag}
but there was no clear signal taking into account the uncertainties of
the measurement.  More recently, the ALICE and STAR Collaborations
reported finite signals~\cite{ALICE:2019aid,Zhou:2019lun}, i.e.,
deviation of $\rho_{00}$ from $1/3$.  Figure~\ref{fig:rho00-lhc} shows
$\rho_{00}$ of $K^{\ast0}$ and $\phi$ mesons as a function of
collision centrality in a form of the number of participants from MC
Glauber simulation, in Pb+Pb collisions at \snn = 2.76 TeV.  At lower
$p_{\rm T}$, the results for both $K^{\ast0}$ and $\phi$ mesons
indicate $\rho_{00}<1/3$.  The STAR results on $\phi$-meson
$\rho_{00}$ show large positive deviation from $1/3$ ($\rho_{00}>1/3$)
for \pt$>1.2$ GeV/$c$ at lower collision energies, while results on
$K^{\ast0}$ are consistent with zero as shown in
Fig.~\ref{fig:rho00-rhic}. The dependence of $\phi$-meson spin
alignment signal on transverse momentum and centrality is not 
systematic; the signal seems to change sign and become negative at
higher transverse momenta as well as in more central collisions, At
present the dependence on transverse momentum and centrality can not
be explained in any scenario.

%\begin{figure}[th]
%%  \centerline{\includegraphics[width=0.65\linewidth]
%  %    {figures/fig_spinalignmentRHIC.pdf}\hspace{0.2cm}}
%  \includegraphics[width=0.49\linewidth]{figures/fig_spinalignmentRHIC.pdf}
%  %\vspace*{0.1cm}
%  \includegraphics[width=0.49\linewidth]{figures/star-rho00-pt.pdf}
%  \centerline{\vspace{0.1cm}\includegraphics[width=0.98\linewidth]
%    {figures/star-rho00-cent.pdf}}
%  \vspace{-0.1cm}
%  \caption{ $K^{\ast0}$ and $\phi$ mesons spin density matrix element
%    with quantization axis along the system orbital angular momentum
%    in Au+Au collisions by the STAR experiment. The results are shown
%    as a function of collision energy \snn, transverse momentum, and
%    centrality.  The figures are adapted from
%    Ref.~\cite{STAR:2022fan}.}
%%  Global spin alignment of $K^{\ast0}$ and $\phi$ mesons shown
%%  as the spin density matrix element $\rho_{00}$ along the system
%%  orbital angular momentum measured in Au+Au collisions as a function
%%  of collision energy \snn by the STAR experiment.  The figure is taken
%%  from Ref.~\cite{STAR:2022fan}.
%\label{fig:rho00-rhic}
%\end{figure}

Note that in the vorticity scenario, the spin alignments signal is
expected to be very small $\Delta\rho \approx (\omega/T)^2 /3 \approx
4 P_H^2/3$.  Taking into account the hyperon global polarization
measurements presented in Fig.~\ref{fig:PHvsRootS-lam}, the spin
alignment signal should be of the order of $10^{-5}$ at the top RHIC
energy and of the order of $10^{-3}$ at lowest BES energy, which is
too small to explain the reported large deviation.
If the vector mesons are produced via quark coalescence, $\rho_{00}$
of vector mesons can be expressed via the quark (antiquark)
polarization $P_{q}(P_{\bar{q}})$ as
$\rho_{00}=(1-P_{q}P_{\bar{q}})/(3+P_{q}P_{\bar{q}}) \approx
(1-4P_{q}P_{\bar{q}}/3)/3$~\cite{Liang:2004xn}.  If
$P_{q}=P_{\bar{q}}=\omega/(2T)$, $\rho_{00}\approx
[1-(\omega/T)^2/3]/3$ which is consistent with the thermal
approach~\cite{Becattini:2016gvu}.
If the particle production for the \pt of interest is dominated by
fragmentation process, the $\rho_{00}$ approximates
$\rho_{00}\approx(1+4P_{q}^2/3)/3$ leading to $\rho_{00} > 1/3$, but
the deviation is again expected to be very small~\cite{Liang:2004xn}.
The only possibility to have large signal in the vorticity based
scenario could arise if the vorticity fluctuations are much larger
than its average. Note that spin alignment signal is proportional to
the (mean root) square of vorticity, while the hyperon polarization is
proportional to its average.

References~\cite{Sheng:2019kmk,Sheng:2020ghv} suggest that the mean
field of $\phi$-mesons could play a role in $\phi$ $\rho_{00}$ but not
in $K^{\ast0}$ $\rho_{00}$ because of mixing of different flavors.
The model involving the strong force seems to explain the energy
dependence of $\phi$ $\rho_{00}$ as shown with the solid line in
Fig.~\ref{fig:rho00-rhic}.  Note that in the given \pt range the
$\phi$ $\rho_{00}$ at the LHC is consistent with zero.

%Also, one of
%the reasons for the larger signal might be due to the fact that the
%$spin alignment measurement probes the magnitude of the polarization
%while the hyperon polarization measurement probes an average of the
%signed-polarization, if the sign of the particle polarization along
%the initial angular momentum depends on azimuthal angle of particles
%as shown in the model calculations~\cite{Becattini_2013,Becattini:2015ska}.

\begin{figure}[th]
  \centerline{\hspace{0.3cm}\includegraphics[width=0.60\linewidth]
    {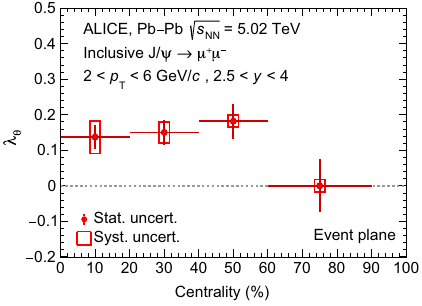}}\vspace{-0.1cm}
\caption{Spin alignment of inclusive $J/\psi$ along the system orbital
  angular momentum in Pb+Pb collisions at \snn = 5.02 TeV from ALICE
  Collaboration~\cite{ALICE:2022dyy}. The measurement was
  performed within $2.5<y<4$ and $2<\pt<6$ GeV/$c$.  }
\label{fig:jpsi-lhc}
\end{figure}

%%A natural question to ask next would be what happens for
%$J/\psi$ meson %which% consists of charm and anti-charm quarks.  Heavy
%quarks are produced via hard scattering of partons at the %beginning
%of the collision, with a time scale of $\tau\sim1/(2m_{\rm
%
Charm quarks are produced via hard scattering of partons at 
the collision, with a time scale of $\tau\sim1/(2m_{\rm HQ})\sim0.1~{\rm fm}/c$.  
Therefore, one may expect larger effect of
the initial magnetic field as well as vorticity on the polarization of 
$J/\psi$ which consists of charm and anti-charm quarks.
The ALICE Collaboration reported inclusive $J/\psi$
polarization relative to the event plane in Pb+Pb collisions at \snn =
5.02 TeV at forward rapidity ($2.5<y<4$)~\cite{ALICE:2022dyy}.
Figure~\ref{fig:jpsi-lhc} shows $J/\psi$ polarization parameter
$\lambda_{\theta}$ (see Eq.~\ref{eq:lambdatheta}) as a function of
centrality.
%where $\lambda_{\theta}$ represents polar anisotropy of
%the decay product distribution in the $J/\psi$ rest frame and relates
%to $\rho_{00}$ as
%$\lambda_{\theta}=(1-3\rho_{00})/(1-\rho_{00})$~\cite{Faccioli:2022peq}.
%Therefore,
Non-zero $\lambda_{\theta}$ means finite polarization of $J/\psi$.
The observed signal of $0.1<\lambda_{\theta}<0.2$ corresponds to
$0.29>\rho_{00}>0.25$, a large negative deviation from 1/3 and
opposite to that of $\phi$-mesons. % (with similar \pt).  Note that this
This measurement was performed at forward rapidity ($2.5<y<4$) and the
measurements at mid-rapidity both at the LHC and RHIC~\cite{Shen:SPIN2023} energies are
needed for a better understanding of the phenomena.  The regeneration
mechanism of the $J/\psi$ production that becomes significant at the
LHC energies, especially at low \pt, might also complicate the
interpretation of the data.

Obviously the spin alignment measurements still need further
investigations, both theoretically and experimentally.  These are
difficult measurements, and as discussed in Sec.~\ref{sec:measure-sa}
and \ref{sec:accspinalign} strongly dependent on complete
understanding of the tracking and acceptance effects.  Future
analyses, in particular based on new high statistics data will
allow us to study the spin alignment in much more detail including
other particles such as charged $D^{\ast}$~\cite{ALICE:2022byg,Micheletti:QM2023} and
$\Upsilon$~\cite{LHCb:2017scf}.

%\clearpage
%---------------------------------------------------------------
\subsection{Polarization along the beam direction}\label{sec:result-Pz}

\begin{figure}[th]
\begin{center}
  \begin{minipage}[b]{0.48\linewidth}
    \includegraphics[width=\columnwidth]{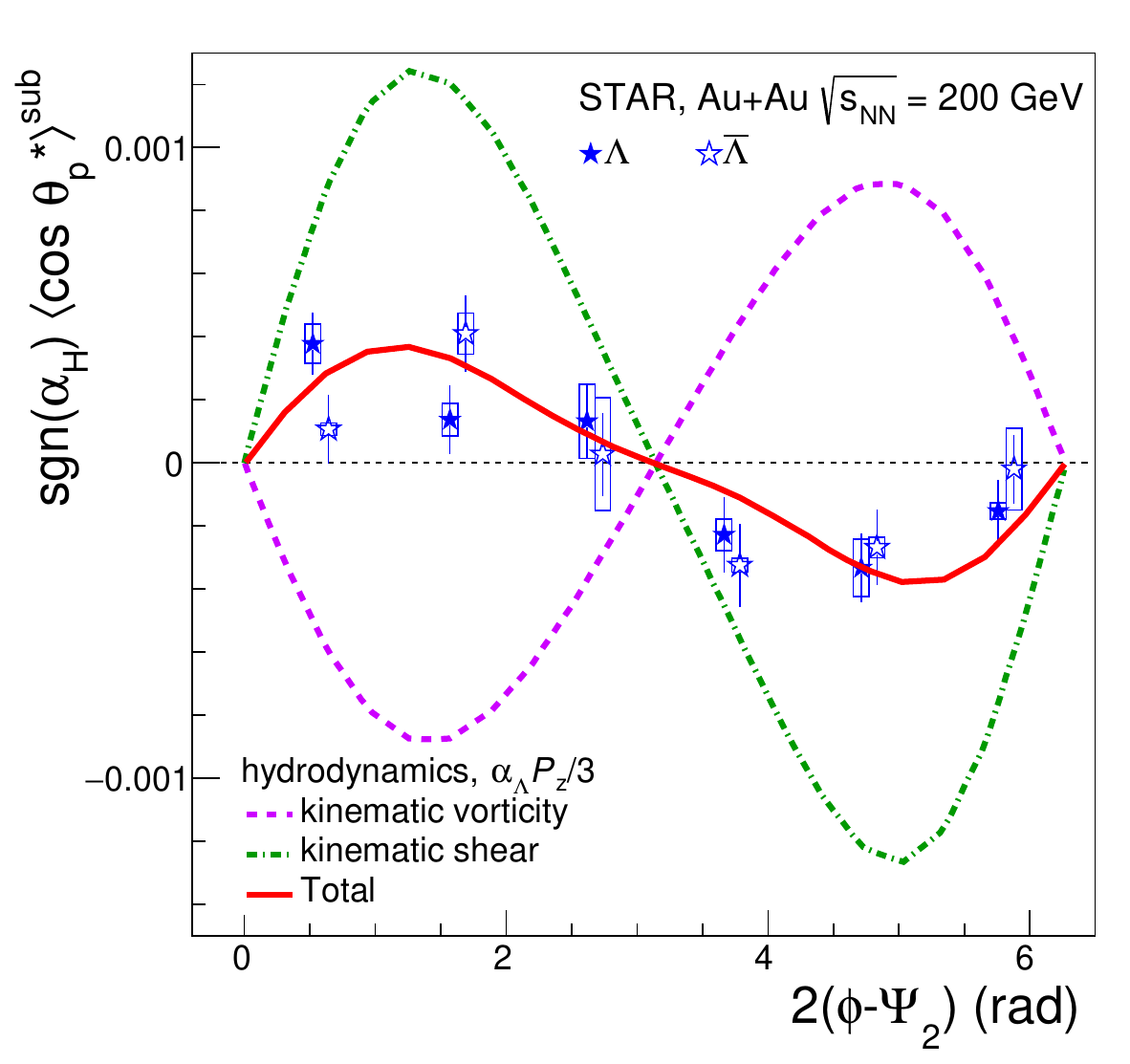}
    \vspace{-6pt}
  \end{minipage}
  \begin{minipage}[b]{0.48\linewidth}
    \includegraphics[width=\columnwidth]{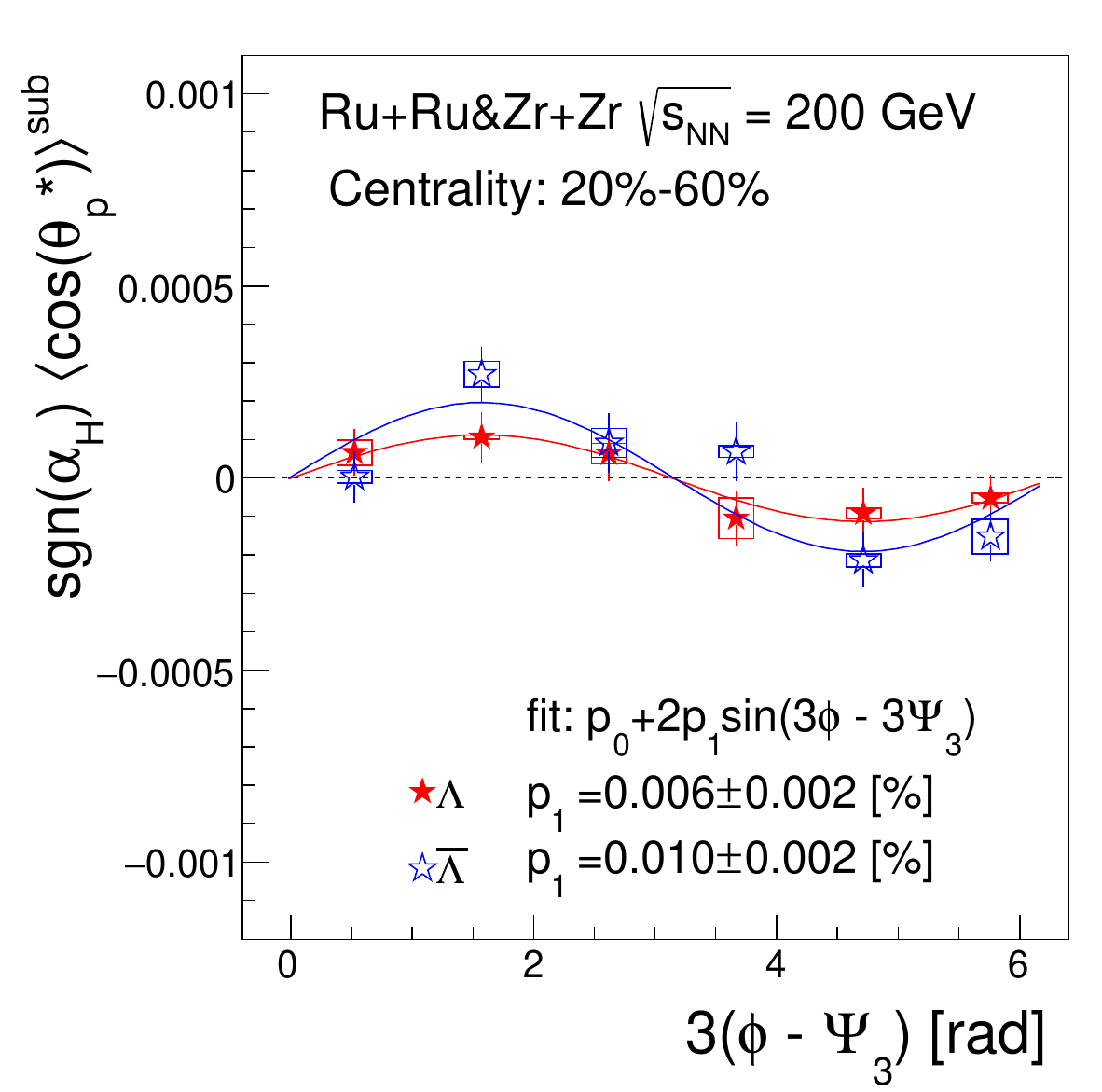}
  \end{minipage}
\caption{ (Left) Raw signal of polarization along the beam direction,
  $\langle\cos\theta_{p}^\ast\rangle$, of \lam and \alam hyperons as a
  function of azimuthal angle relative to the second-order event plane in
  Au+Au collisions at \snn = 200~GeV~\cite{Adam:2019srw}.  Hydrodynamic
  model calculations including kinematic vorticity and kinematic shear
  separately, as well as the sum of the two are shown by
  lines~\cite{Becattini:2021iol}.  (Right) Same as left figure for
  polarization relative to the third-order event plane in R+Ru and Zr+Zr
  collisions at \snn =~200 GeV~\cite{STAR:2023eck}. }
\label{fig:rawpz}
\end{center}
\end{figure}

\begin{figure}[th]
\vspace{-0.2cm}
\centerline{\includegraphics[width=0.68\linewidth]{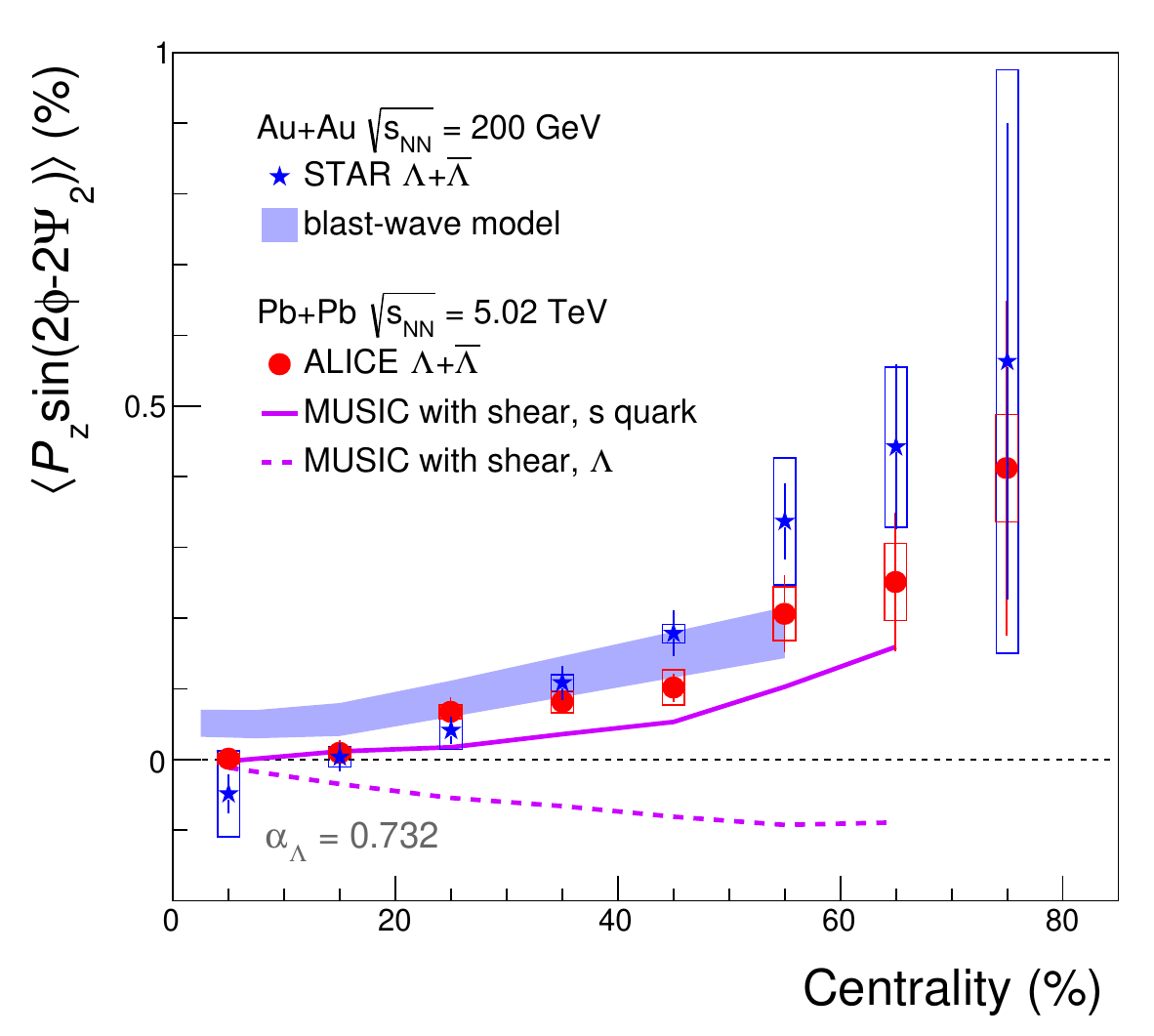}}
\vspace{-0.1cm}
\caption{The second-order sine modulation of \lam polarization along the
  beam direction as a function of centrality at RHIC~\cite{Adam:2019srw} and
  the LHC~\cite{ALICE:2021pzu} compared to various model calculations~\cite{Adam:2019srw,Fu:2021pok}. 
  The experimental data are rescaled with \lam decay parameter
  $\alpha_{\Lambda}=0.732$~\cite{Zyla:2020zbs}.}
\label{fig:pzcent}
\end{figure}

%In noncentral heavy-ion collisions, the system exhibits stronger expansion in
%the transverse plane in direction of the the impact parameter (a shorter
%axis of the elliptically shaped reaction region, see
%Fig.~\ref{fig:BW23}(left)) than in the direction perpendicular to it, the
%phenomenon known as elliptic flow. Such
%(see also the discussion in
As discussed in Sec.~\ref{sec:ani-flow}, anisotropic transverse
flow leads to nonzero vorticity component along the beam direction,
with the direction of vorticity changing with the azimuthal
angle~\cite{Voloshin:2017kqp,Becattini:2017gcx}, as depicted by open
arrows in Fig.~\ref{fig:BW23}(left). The polarization along the beam
direction $P_{\rm z}$ was first measured with \lam hyperons by the
STAR Collaboration at RHIC~\cite{Adam:2019srw}.  Later, it was also
observed by the ALICE Collaboration at the LHC
energy~\cite{ALICE:2021pzu}.  As expected from the elliptic flow
picture, $P_{\rm z}$ exhibits a quadrupole or $\sin(2\phi)$ pattern as
shown in Fig.~\ref{fig:rawpz}(left).  The polarization is quantified
by a second-order Fourier sine coefficient and studied as a function
of centrality, see Fig.~\ref{fig:pzcent}.  The results show a clear
centrality dependence similar to that of elliptic flow except in most
peripheral collisions.  The polarization magnitudes at RHIC and the
LHC are rather similar, indicating weak collision energy dependence
unlike in the global polarization case.

It was found that hydrodynamic and transport models that successfully
reproduce the energy dependence of the global polarization fail badly
in predictions of the magnitude and the sign (phase) of the azimuthal
angle modulation
differently~\cite{Becattini:2017gcx,Xia:2018tes,Fu:2020oxj,
  Florkowski:2019voj,Sun:2018bjl,Xie:2019jun,Adam:2019srw,Wu:2019eyi}.
This was true for several hydrodynamics models using different
approaches and initial conditions.  The chiral kinetic approach
accounting for the nonequilibrium effects of the spin degrees of freedom
gives the correct sign of the $P_{\rm z}$
modulation~\cite{Sun:2018bjl}.  Interestingly, the Blast-Wave model
which is a simplified model of hydrodynamics with a few freeze-out
parameters~\cite{STAR:2004qya} (taken from STAR publication in 2005!)
describes the data very well~\cite{Adam:2019srw}.

More recently, two independent groups pointed out
that accounting for contribution from the fluid velocity
shear (see Sec.~\ref{sec:SIP-SHE}) 
%should also contribute to the spin polarization (see
%Sec.~\ref{sec:intro-local}) and
might help to explain the disagreement between the data and
theoretical calculations.  As shown in Fig.~\ref{fig:rawpz}(left), the
contribution from the kinematic shear, as that in the hydrodynamic
model~\cite{Becattini:2021iol}, exhibits an opposite sign in $P_{\rm
  z}$ modulation to that of the kinematic vorticity, and as a
consequence, combining the two effects leads to a trend similar to the
data if additionally the model assumes the isothermal freeze-out.
Hydrodynamic model (MUSIC with AMPT initial conditions) including the
shear contribution~\cite{Fu:2021pok}, and assuming that \lam inherits
the polarization from the strange quark, can also qualitatively
describe the measurements including the centrality dependence as shown
in Fig.~\ref{fig:pzcent}. But the predictions change to the opposite
sign if the polarization is calculated using \lam mass.  It should be
noted that the thermal vorticity and shear contributions are largely
canceled
out~\cite{Florkowski:2021xvy,Yi:2021ryh,Sun:2021nsg,Alzhrani:2022dpi}
and the final result depends strongly on the detailed implementation
of those contributions.  Thus the spin sign puzzle still needs more
investigations.

\begin{figure}[bth]
\includegraphics[width=0.49\linewidth]{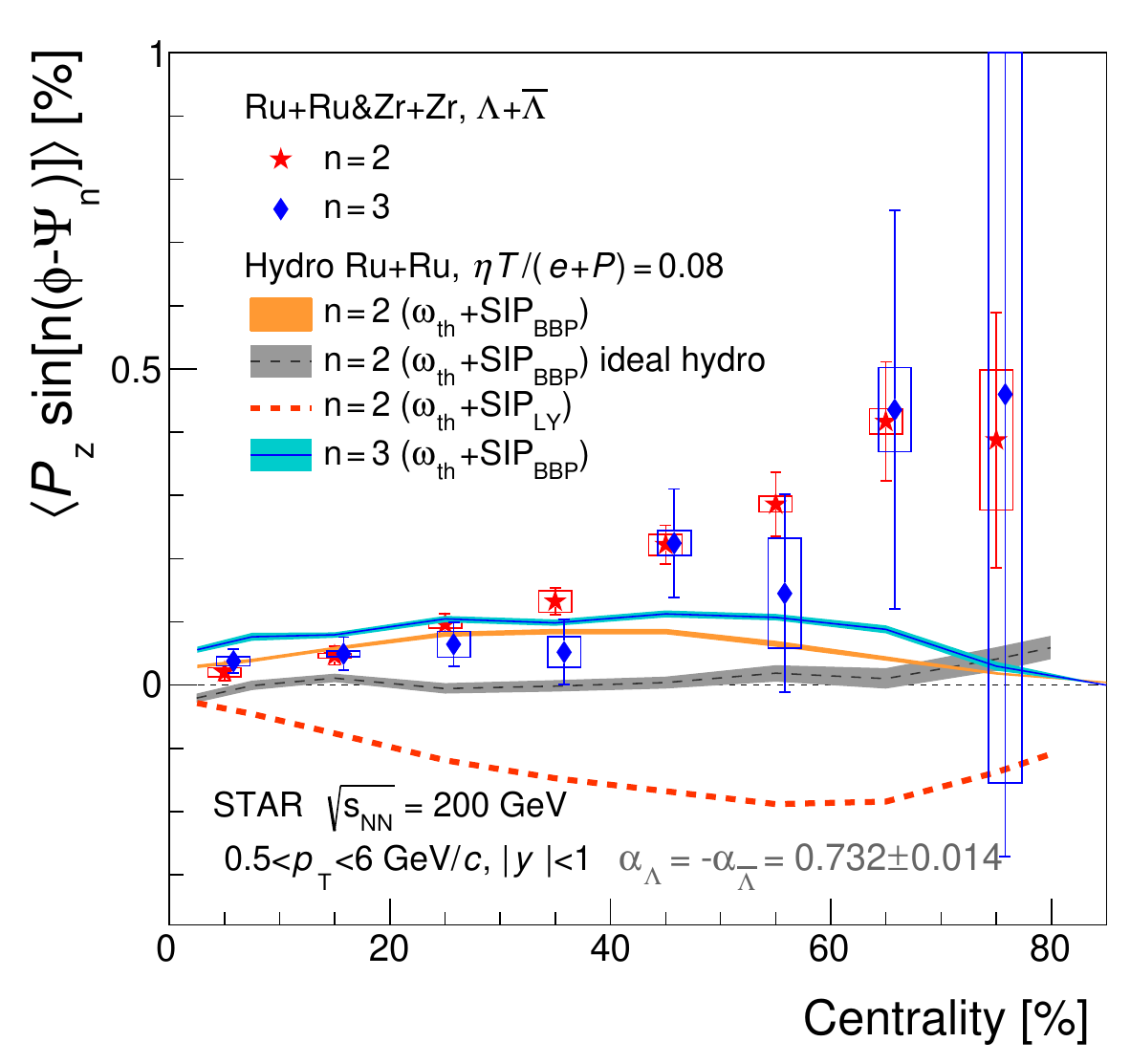}
\includegraphics[width=0.49\linewidth]{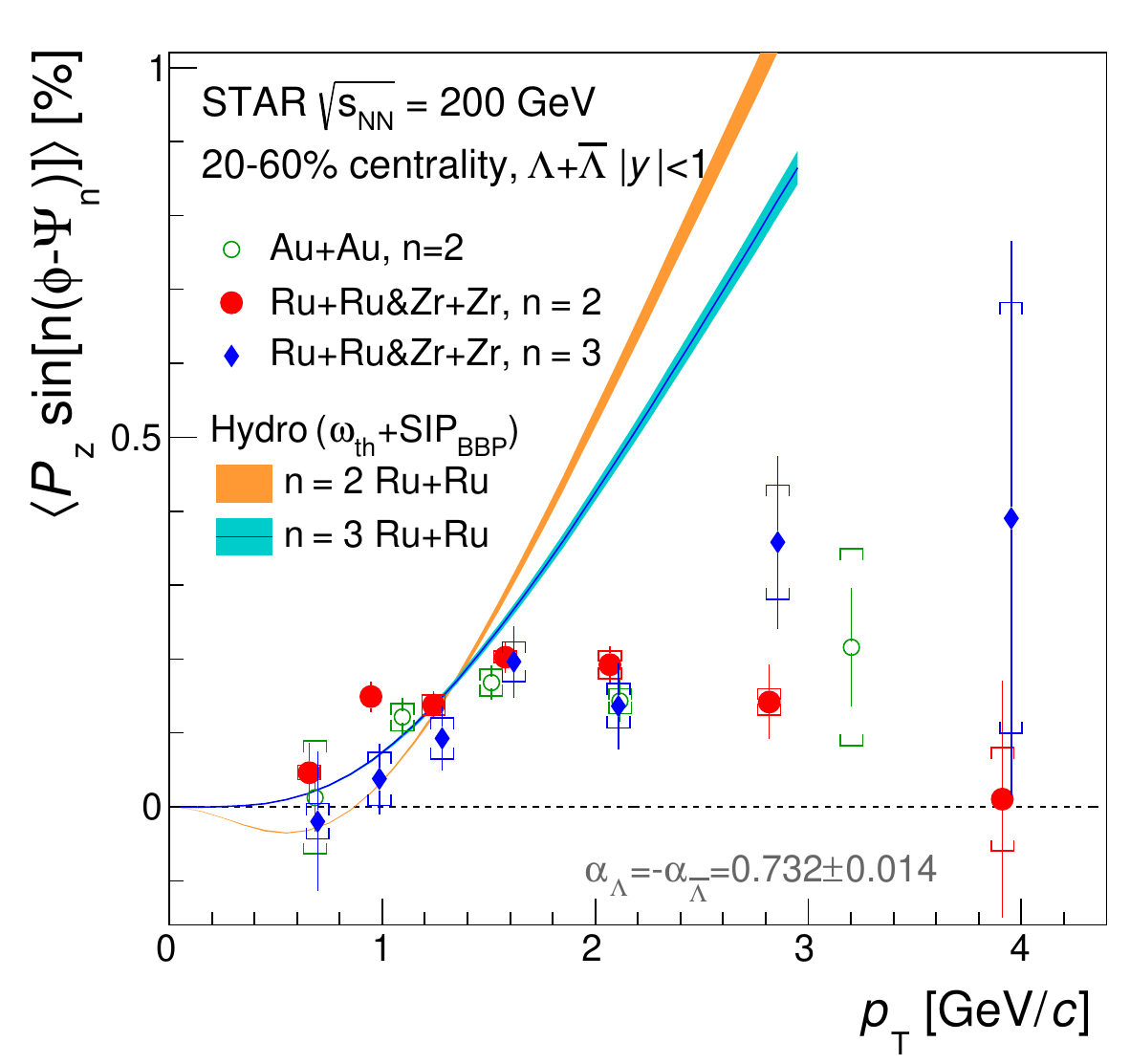}
\caption{The second and third-order sine modulation of \lam polarization
  along the beam direction as a function of centrality (left) and hyperons'
  transverse momentum (right) in Ru+Ru and Zr+Zr collisions at \snn =
  200~GeV~\cite{STAR:2023eck}. Solid and dashed lines are the calculations
  from hydrodynamic model with particular implementation of the shear
  induced polarization (SIP)~\cite{Alzhrani:2022dpi}.  See texts for the
  detail.  }
\label{fig:PzNiso}
\end{figure}

As predicted in Ref.~\cite{Voloshin:2017kqp},  higher harmonic
anisotropic flow should also lead to a similar
vorticity structure and polarization along the beam
direction.  Recently, the STAR Collaboration has
reported \lam polarization along the beam direction relative
to the third harmonic event plane in isobar Ru+Ru and Zr+Zr collisions
\snn = 200~GeV~\cite{STAR:2023eck}; these results are shown in
Fig.~\ref{fig:rawpz}(right). The sine modulation of $P_{\rm z}$
relative to the third-order event plane was observed similarly to the
second-order case, indicating a sextupole pattern of vorticity induced
by triangular flow as depicted in Fig.~\ref{fig:BW23}(right).

Figure~\ref{fig:PzNiso}(left) shows $P_z$ sine coefficients relative
to the second and third order harmonic event planes as a function of
centrality in the isobar collisions.  The third-order result seems to increase towards
peripheral collisions as the second-order does.  Calculations from
hydrodynamic model with two different implementations of the shear
induced polarization (SIP), based on Ref.~\cite{Becattini:2021suc} by
Becattini-Buzzegoli-Palermo (BBP) and on Ref.~\cite{Liu:2021uhn} by
Liu-Yin (LY), are also compared. The calculations with ``SIP$_{\rm
  BBP}$" reasonably well describe the data for both the second and
third-orders except peripheral collisions. The calculation with
``SIP$_{\rm LY}$" leads to the opposite sign to the data but note that
it provides the correct sign if the mass of strange quark is used
instead of \lam mass as shown in Fig.~\ref{fig:pzcent}.  It is also
worth to mention that the calculation with a nearly zero specific
shear viscosity (denoted as ``ideal hydro") leads to almost zero $P_z$
sine coefficient, which indicates that the $P_z$ measurement could
provide an additional constraint on the shear viscosity of the medium.
Figure~\ref{fig:PzNiso}(right) shows \pt dependence of the second and
third-order $P_{\rm z}$ sine coefficients.  The third-order result is
found to be comparable in magnitude to the second-order result,
slightly smaller at low \pt and showing a hint of overpassing the
second-order at high \pt. This trend is similar to what was observed
in \pt dependence of the elliptic and triangular
flow~\cite{PHENIX:2014uik}, which further supports the picture of
anisotropic-flow-driven polarization.  The model incorporating the
shear induced polarization of SIP$_{\rm BBP}$ is comparable to the
data at low \pt but not the \pt dependence in detail.

\section{Open questions  and future perspective}
%==============  open questions and perspective ==========

Summarizing the discussion in Sec.~\ref{sec:result}, one tend to
conclude that while the theoretical description of the global
polarization, including its energy dependence, is rather good, our
understanding of the local polarization measurements, in particular
the azimuthal angle dependence of the polarization along the beam
direction is far from satisfactory. Surprisingly, the data is much
better described by ``naive'' Blast-Wave model including only
nonrelativistic vorticity, than by more sophisticated hydrodynamical
calculations (including contribution from temperature gradients and
acceleration), which very often differ from the data at the
qualitative level. Recent calculations including the shear induced
polarization make the comparison somewhat better, but still
unsatisfactory.  The disagreements with theoretical models definitely
need further investigation in future, including the role of different
freeze-out scenarios, validity of Cooper-Frye prescription, relative
contributions of kinematic vorticity, acceleration, SIP, SHE, and
temperature gradients.  Comparison of more advanced calculations with
new measurements should be also able to provide information of
vorticity evolution and spin equilibration relaxation times.

From experimental point of view, in the next few years several new
precise measurements will be performed to shed more light on the
topics of interest, such as particle-antiparticle polarization
splitting, rapidity and azimuthal angle dependencies, and particle species
dependence.  Below we list possible near-future measurements intended
to provide more information on the vorticity and polarization
phenomena in heavy-ion collisions.
\begin{itemize}
\item Polarization splitting between particles and antiparticles,
  including particles with larger magnitude of the magnetic moment
  such as $\Omega$. It will further constrain the magnetic field time
  evolution and its strength at freeze-out, and the electric
  conductivity of quark-gluon plasma.
\item Precise measurements of multistrange hyperon polarization to
  study particle species dependence and confirm the vorticity-based
  picture of polarization. Measurement with $\Omega$ will also
  constrain unknown decay parameter $\gamma_{\Omega}$.
\item Precise differential measurements of the azimuthal angle and
  rapidity dependence of $P_J~(\pmy)$.
\item Detailed measurement of $\pz$ induced by elliptic and higher
  harmonic flow. In particular this study could help to identify the
  contribution from SIP, which is expected to be different for
  different harmonics.
\item Application of the event-shape-engineering
  technique~\cite{Schukraft:2012ah} testing the relationship between
  anisotropic flow and polarization.
\item Measuring $\px$ to complete all the components of polarization
  and compare the data to the Glauber estimates and full
  hydrodynamical calculations.
\item Circular polarization $P_{\phi}$ to search for toroidal vortex structures
\item The particle-antiparticle difference in the polarization
  dependence on azimuthal angle at lower collision energies testing the
  Spin-Hall Effect.
\item Understanding of the vector meson spin alignment measurements including
  new results with corrections of different detector effects.
\item Measurement of the hyperon polarization correlations to access
  the scale of vorticity fluctuations.  
\item Measurement of the hyperon polarization in $pp$ collisions to
  establish/disprove possible relation to the single spin asymmetry effect. 
\end{itemize}
%

%==================================================================

\section{Summary}

The polarization phenomena in heavy-ion collisions appeared to be an
extremely interesting and important subject overarching such questions
as the nature of the spin and spin structure of the hadrons, evolution
of the QGP and its hadronization, and finally the freeze-out of the
system. While many, or better to say, most of the details of the
entire picture is far from being even well formulated, it is clear
that following this direction we might expect many important
discoveries.

The observed global polarization of hyperons in heavy-ion collisions
is found to be well described by hydrodynamic and microscopic
transport models based on the local vorticity of the fluid averaged
over the freeze-out hypersurface under assumption of the local thermal
equilibrium of the spin degrees of freedom. Furthermore, the
measurement of hyperon polarization along the beam direction confirmed
the local vorticity induced by anisotropic collective flow, adding to
the evidences of ideal fluid dynamics of the quark-gluon plasma.
These measurements opened new direction to study the dynamics of
quark-gluon plasma and spin transport in the hot and dense medium,
triggering a lot of theoretical interest on spin dynamics in general.
Despite the successful description of the average global polarization,
when looking into the detailed comparison between the data and models
in differential measurements, there are still many open questions to
be solved. The spin alignment measurements of vector mesons are very
intriguing, but far from satisfactory understanding.  More precise
measurements with different particle species and a wider detector
acceptance that will be available in near future at RHIC and the LHC
and future experiments at new facilities will be extremely helpful to
shed light on existing issues.

%==================================================================
\section*{Acknowledgments}
%This section should come before the References. Funding
%information may also be included here.

The discussions with F. Becattini,
%M. Lisa\red{\bf ?}, J. Schukraft\red{\bf ?},
and C. Shen
are gratefully acknowledged. 

S. Voloshin is supported by the U.S. Department of Energy Office of
Science, Office of Nuclear Physics under Award No. DE-FG02-92ER40713.
T. Niida is supported by JSPS KAKENHI Grant Number JP22K03648.

%=======================================================

%\bibliographystyle{ws-ijmpe}
\bibliographystyle{utphys}
\bibliography{ref_ijmpepol}

%========================================================
%\appendix
%
%\section{Appendices}
%
%Appendices should be used only when absolutely necessary. They
%should come after the References. If there is more than one
%appendix, number them alphabetically. Number displayed equations
%occurring in the Appendix in this way, e.g.~(\ref{a1}), (A.2),
%etc.
%

%=============================================================
%\appendix

\end{document}